\documentclass[prb,twocolumn,floatfix,showpacs,superscriptaddress]{revtex4}
\usepackage{amsmath}
\usepackage[dvips]{graphicx}
\usepackage{natbib}
\usepackage{subfigure}

\begin{document}

\title{Mott physics and first-order transition between two metals in the normal state phase diagram of the two-dimensional Hubbard model}
\author{G. Sordi}
\affiliation{D\'epartement de physique and Regroupement qu\'eb\'equois sur les mat\'eriaux de pointe, Universit\'e de Sherbrooke, Sherbrooke, Qu\'ebec, Canada J1K 2R1}
\affiliation{Theory Group, Institut Laue Langevin, 6 rue Jules Horowitz, 38042 Grenoble Cedex, France}
\author{K. Haule}
\affiliation{Department of Physics \& Astronomy, Rutgers University, Piscataway, NJ 08854-8019, USA}
\author{A.-M. S. Tremblay}
\affiliation{D\'epartement de physique and Regroupement qu\'eb\'equois sur les mat\'eriaux de pointe, Universit\'e de Sherbrooke, Sherbrooke, Qu\'ebec, Canada J1K 2R1}
\affiliation{Canadian Institute for Advanced Research, Toronto, Ontario, Canada, M5G 1Z8}

\date{\today}

\begin{abstract}
For doped two-dimensional Mott insulators in their normal state, the challenge is to understand the evolution from a conventional metal at high doping to a strongly correlated metal near the Mott insulator at zero doping.
To this end, we solve the cellular dynamical mean-field equations for the two-dimensional Hubbard model using a plaquette as the reference quantum impurity model and  continuous-time quantum Monte Carlo method as impurity solver. 
The normal-state phase diagram as a function of interaction strength $U$, temperature $T$, and filling $n$ shows that, upon increasing $n$ towards the Mott insulator, there is a surface of first-order transition between two metals at nonzero doping.
That surface ends at a finite temperature critical line originating at the half-filled Mott critical point.
Associated with this transition, there is a maximum in scattering rate as well as thermodynamic signatures.
These findings suggest a new scenario for the normal-state phase diagram of the high temperature superconductors.
The criticality surmised in these systems can originate not from a T=0 quantum critical point, nor from the proximity of a long-range ordered phase, but from a low temperature transition between two types of metals at finite doping.
The influence of Mott physics therefore extends well beyond half-filling.
\end{abstract}
\maketitle


\section{Introduction}

The discovery of quantum oscillations in layered high-temperature superconductors~\cite{nicolas,leboeuf,jaudet,sebastian,yelland,bangura} has brought renewed attention to the normal phase below the superconducting dome.
The challenge posed is to understand how these systems evolve as a function of carrier concentration in the absence of the superconducting phase~\cite{LouisREV}.
At large doping, the copper-oxide layers are metallic.
On the other hand, at zero doping these systems are Mott insulators~\cite{andersonSCIENCE,andersonBOOK}, i.e. systems that are insulating due to the strong screened Coulomb interactions between the electrons.
Thus, as a function of the carrier concentration, a Mott metal-insulator transition (MIT) occurs, i.e. a transition driven by the on-site Coulomb repulsion between the electrons~\cite{mott,ift}. We are challenged to understand the Mott transition driven by carrier concentration in the normal phase~\cite{lee}.

At the theoretical level, to address this physics we need to study the competition between the kinetic band effects, that delocalize the electrons in the lattice, and the effects due to screened interactions that localize them.
These two ingredients are present in the Hubbard model, which contains a kinetic energy term that describes the band structure, plus an interaction term that represents the screened local Coulomb repulsion between electrons occupying the same site of the lattice.
More than fifty years of intense research shows that this simple model has a surprisingly rich set of solutions that may help capture the complex behavior observed in nature with minimal assumptions.

Even the simple-looking Hubbard model poses a formidable challenge to theory. Dynamical mean-field theory (DMFT) has played a central role in advancing our knowledge of the physics contained in the Hubbard model and has proven to be a useful method to investigate the Mott metal insulator transition. In essence, dynamical-mean-field theory embeds a single site~\cite{rmp,mv} or a cluster~\cite{maier,kotliarRMP,tremblayR} in an infinite self-consistent bath of non-interacting electrons.
The latter problem is then amenable to accurate numerical solutions. 
This approach can also be justified from a variational perspective~\cite{Potthoff:2003b}.
DMFT with a single site immersed in a bath is exact in infinite dimension~\cite{rmp,mv} and provides a quite accurate mean field description of materials~\cite{kotliarRMP} and of the Mott transition~\cite{limelette} in three dimensions. 
The picture is that the Mott insulator can be driven towards a correlated metallic state through a first-order transition by tuning correlation strength, or temperature, or doping~\cite{rmp,mzk,mkz,mck,antoine,sahana,lange,gabiEPJB}.
Here the fundamental unifying concept is the first-order transition.
That transition dominates the entire normal phase diagram of the model, and is relevant at finite temperature even if the actual Mott transition is hidden by the onset of some long-range order.

Something similar occurs in the Fermi liquid description of the normal state of weakly correlated materials. 
There, the zero-temperature Fermi liquid fixed point can be masked by a long-range ordered state that is the true ground state. 
Nevertheless, as long as one is not too close to the transition to long-range order, the normal state is well described by the finite temperature Fermi liquid that emerges from the zero-temperature Fermi liquid fixed point. 
While the Fermi liquid provides a mean-field description of a phase where the self-energy is analytical $(Im\Sigma \propto \omega^2$), DMFT also allows a phase, such as the Mott insulator, where the self-energy is not an analytical function of frequency~\cite{PhillipsRMP:2010}.

For strongly-interacting layered materials, such as the above-mentioned cuprate high temperature superconductors, or for the layered organic conductors, the description provided by the single-site DMFT solution of strongly correlated electron models is valid only at very large doping or at temperatures that are in fact much too large to be interesting. 
This is because single-site DMFT neglects spatial fluctuations, and for these systems the short-range correlations play an important role due to their low dimensionality. 
For example, the observed pressure-driven first-order Mott transition in organic superconductors of the BEDT family~\cite{Lefebvre:2000,LimeletteBEDT:2003,kagawa1,kagawa2} has the wrong slope in the pressure-temperature plane when compared with the single-site DMFT solution. This is corrected~\cite{ohashi:2008,Liebsch:2009} by cluster extensions of DMFT~\cite{Hettler:1998,lkAF,gabiCDMFT,maier,kotliarRMP,tremblayR} that incorporate short-range magnetic correlations in the theoretical description.
Similarly, quantum Monte Carlo calculations on the square lattice at half-filling show that a Mott gap opens up at finite temperature independently of antiferromagnetism starting around~\cite{Vekic:1993} $U\approx 6t$, with $U$ the interaction strength and $t$ nearest-neighbor hopping. It is only with extensions of DMFT on a plaquette that this result is recovered~\cite{phk,gullEPL,balzer}.  We stress that as larger and larger cluster sizes are considered~\cite{werner8}, the description of the normal state will become more and more accurate at lower temperature, but it will also fail when long-range order sets in. Hence, we argue that calculations, such as ours, that are based on a plaquette and reproduce the $U\approx 6t$ result mentioned above are an accurate mean-field description of the normal sate. Hence, this is the type of approach that we adopt in the present paper.

Cluster extensions of DMFT are however computationally more expensive than their single-site counterpart. Nevertheless,
useful physical insights on the phase diagram of the Hubbard model have been obtained within that framework and have helped elucidate some important properties of the copper oxide superconductors.
Among the main results, we mention pseudogap formation at low doping~\cite{kyung,st,jarrellNFL,liebsch}, the presence of antiferromagnetism and superconductivity~\cite{lkAF,maierAF,kancharla,davidAF,massimoAF,markus, civelli1,civelli2,hauleDOPING}, and the momentum dependence of the electronic properties along the Fermi surface~\cite{olivierMott,civelliBreakup,st,stanescuK,sakaiPRL,sakaiPRB}.  We caution that cluster extensions of DMFT come in two varieties: The Dynamical Cluster Approximation~\cite{Hettler:1998} (DCA) and cellular DMFT~\cite{gabiCDMFT,lkAF}. For simplicity we did not distinguish between both approaches in the previous discussions. They generally give qualitatively similar results. 
Here we shall employ cellular DMFT.

Returning to the normal state, the overall picture for the Mott transition in cluster extensions of DMFT have been lacking because cluster methods, as already mentioned, are computationally expensive.
However, a few years ago there was a breakthrough. 
The continuous-time quantum Monte Carlo technique in the hybridization expansion~\cite{Werner:2006,WernerCTQMC,hauleCTQMC,millisRMP} now provides us with the possibility to explore the full phase diagram of the model within the cluster DMFT method.
This is because this algorithm considerably reduces the computational time, allowing both unprecedented accuracy and studies over a wider range of parameters.
Recent investigations using this method and other continuous-time quantum Monte Carlo techniques~\cite{Rubtsov:2003}, have determined the interaction driven Mott MIT, revealing sharp modifications to the single-site picture~\cite{phk,gullEPL,werner8}.
The transition driven by carrier concentration, more relevant for the high-temperature superconductors, is also currently under intense investigation~\cite{hauleDOPING,gullEPL,werner8,gullPRB,michelEPL,michelPRB,jarrell:2009,khatami,michelCFR,sakaiPRB}.
Motivated by the physics of the cuprates, almost all the studies focus on the large interaction regime where the Mott gap is well developed and they consider the effects of different band structure parameters in order to capture the striking particle-hole asymmetry observed in those compounds.
This direction of research produced several important results, however a unified scenario for the Mott transition realized in cluster extensions of DMFT for the Hubbard model was still missing.

In our recent work~\cite{sht} and in the present study, we take a different approach to address this problem.
Instead of focusing on a region of the parameter space where the interaction strength is large, we map out the {\it whole} normal state phase diagram of the two-dimensional Hubbard model as a function of interaction, temperature and doping. To capture the general features of the localization-delocalization Mott physics, we simply consider nearest-neighbor hopping.

Our strategy is motivated by one of the key lessons of the single-site DMFT solution of the Hubbard model: a first-order transition governs the entire phase diagram of the model.
We ask whether the Mott transition still dominates the phase diagram and how it is modified by the short-range correlations taken into account through cellular DMFT.
As mentioned above, in the Hubbard model the Mott transition can be tuned by temperature, by the interaction strength (interaction driven MIT, relevant for the layered organics) and by the carrier concentration (doping driven MIT, relevant for the cuprates). Fundamentally, one is exploring the same strong coupling physics from different perspectives.
At half-filling, the cluster DMFT solutions of the model {\it do} display a first-order transition, as a function of the interaction strength, between a correlated metallic state and a Mott insulator~\cite{phk,gullEPL,balzer}.

What is the fate of this first-order transition as a function of the carrier concentration?
As reported in our recent work~\cite{sht} and in great detail here, the full mapping of the normal-state phase diagram reveals that upon increasing the carrier concentration towards the Mott insulator, there is indeed a surface of first-order transition. This is distinct from the surface of first-order transition found at fixed $U$ as a function of second-neighbor hopping in earlier work~\cite{khatami,mjm}.
Thus, as in the single-site case, in cellular DMFT the finite temperature Mott transition provides the key to understand the whole phase diagram.
In addition, contrary both to single-site DMFT and to conventional wisdom, the first-order transition, emerging at the half-filled Mott critical endpoint, progressively moves to large doping as the interaction strength increases. Hence the transition can occur between two metallic states, one of which evolves continuously from the Mott insulator, a surprising effect brought about by the short-range correlations.

The first-order transition and the associated critical line that we find are not only new, they also provide a unified picture for a host of previously known results.
For example, the first-order transition was not detected at large values of the interaction by previous works most likely because the critical line moves rapidly to lower temperature with increasing interaction strength, falling below the temperatures accessible to date.
Hence, previously reported features of the phase diagram of the Hubbard model~\cite{gullPRB,michelEPL,michelPRB,jarrell:2009,khatami,michelCFR,sakaiPRB,liebsch}, whether it is thermodynamic properties, scattering rate, momentum differentiation or other, appear in a different light. 
As we shall see, our analysis allows one to identify these features as precursors of the first-order transition, hence calling for their re-evaluation. It is in this decisive sense that our work goes beyond -and is compatible with- previous studies.

Compared to our recent letter~\cite{sht},
the present work extends and provides further results on the thermodynamics, on the spin susceptibility, on results for other values of the parameters and many additional details and a refined interpretation.
In Section~\ref{Sec:Methodology} we briefly discuss the model and method.
The peculiar normal-state phase diagram of the two-dimensional Hubbard model is the topic of Section~\ref{Sec:PhaseDiagram}.
In Section~\ref{Sec:Transition} we discuss the first-order transition that controls the physics of the phase diagram.
Section~\ref{Sec:Thermodynamics} contains the thermodynamic properties of the phases separated by the transition.
The characterization of these phases continues in Section~\ref{Sec:Phases} where a signature of critical behavior through a large scattering rate is found near the transition.
We identify the physical origin of this critical behavior in the Section~\ref{Sec:Plaquette}.
In Section~\ref{Sec:Discussion} we reexamine the phase diagram and propose a possible new scenario for the surmised criticality in high-temperature superconductors. It can originate not from a quantum critical point, nor from the proximity of a long-range ordered phase, but from a finite-temperature transition at finite doping coming from the influence of Mott physics well beyond half-filling.

\section{Model and method}
\label{Sec:Methodology}

The Hamiltonian for the two-dimensional Hubbard model on a square lattice reads
\begin{equation}
  H = -\sum_{ij\sigma} t_{ij} c_{i\sigma}^\dagger c_{j\sigma}
  + U \sum_{i}  \left(n_{i \uparrow }-\frac{1}{2}\right) \left(n_{i \downarrow }-\frac{1}{2}\right)
  - \mu\sum_{i\sigma} n_{i\sigma}.
\label{eq:HM}
\end{equation}
Here $c_{i\sigma}$ and $c^+_{i\sigma}$ operators annihilate and create electrons on site $i$ with spin $\sigma$, and $n_{i\sigma}=c^+_{i\sigma}c_{i\sigma}$ is the number operator.
In this article we focus on the physics arising from the proximity to a Mott insulating state, so we use a simple hopping amplitude $t_{ij}$ between nearest neighbors only, so that the bare dispersion, obtained from the Fourier transform of $t_{ij}$, is given by $\epsilon(k) =-2t [ \cos(k_x)+\cos(k_y) ]$.
$U$ is the energy cost of double occupation at each site of the lattice,
$\mu$ is the chemical potential which, for a given value of $U$, controls the occupation $n=1/N \sum_{i\sigma} \langle n_{i\sigma} \rangle$ where $N$ is the number of sites.

We solve this model using cellular dynamical mean-field theory \cite{gabiCDMFT,maier,kotliarRMP}.
This approach maps the lattice problem Eq.~(\ref{eq:HM}) onto a quantum impurity problem consisting of a cluster of sites, here a $2\times2$ plaquette, embedded in a bath that is determined self-consistently in such a way that infinite lattice and plaquette have the same self-energy. 
The action of the quantum impurity problem (plaquette coupled to the bath) is given by \cite{kotliarRMP}
\begin{equation}
  S = S_{c} +\int_{0}^{\beta} d\tau \int_{0}^{\beta} d\tau' \sum_{R R'} \psi_{R}^{\dag}(\tau) \hat{\Delta}_{R R'}(\tau,\tau') \psi_{R'}(\tau') ,
\label{eq:action}
\end{equation}
where $S_{c}$ is the action of the cluster, ${R}$ labels the cluster sites, and $\hat{\Delta}$ is the $4\times4$ bath hybridization matrix.
The self-consistency condition that fixes the bath hybridization matrix $\hat{\Delta}$ is
\begin{equation}
\begin{split}
\hat{\Delta}(i\omega_{n}) = & \, i\omega_{n} +\mu -\hat{t}_{c}-\hat{\Sigma}_{c}(i\omega_{n})  \\
                      & \, -\left[ \sum_{\tilde{k}} \frac{1}{i\omega_{n} +\mu -\hat{t}(\tilde{k}) -\hat{\Sigma}_{c}(i\omega_{n})}\right] ^{-1},
\end{split}
\label{eq:SCC}
\end{equation}
where $\hat{\Sigma}_{c}$ is the cluster self-energy matrix, $\hat{t_c}$ is the plaquette hopping matrix, $\hat{t}(\tilde{k})$ is the lattice matrix of hopping in the supercell notation and $\tilde{k}$ runs over the reduced Brillouin zone of the superlattice.

There are a variety of techniques to solve the quantum impurity problem Eq.~(\ref{eq:action}) \cite{kotliarRMP,maier,tremblayR}.
Here we use the recently developed continuous-time quantum Monte Carlo method \cite{Werner:2006, WernerCTQMC,hauleCTQMC}.
This approach relies on the Monte Carlo summation of all diagrams generated by the expansion of the cluster-bath (impurity) action Eq.~(\ref{eq:action}) with respect to the hybridization $\hat{\Delta}$.
Ref.~\onlinecite{hauleCTQMC} contains the details of the present implementation.

The CTQMC method is a powerful and perfectly adapted technique to map out the phase diagram of the Hubbard model, because it permits to readily access all regions of doping, from large doping down to the Mott insulating state, the intermediate to low temperature regime that have so far proven inaccessible with usual Hirsch-Fye quantum Monte Carlo~\cite{hf} for instance, and a large range of $U$ with high efficiency~\cite{gullCOMP}.
In addition, this method is statistically exact and does not have errors associated with time discretization or bath parametrization, and therefore we can determine the phase diagram of the Hubbard model with unparalleled accuracy.

To obtain high quality data, required to determine the phases of the system, we typically perform $5\times10^6$ Monte Carlo sweeps per processor, averaged over 40 processors.
When necessary, for example close to phase boundaries, we may do up to $10^7$ sweeps per processor averaged on 64 processors.
Cellular DMFT is an iterative approach, and in generic regions of parameter space we have studied the cluster observables converge in less then $20-30$ iterations, but hundreds may be necessary close to phase boundaries.

In the actual CTQMC implementation, it is useful to write cluster quantities, such as the hybridization $\hat{\Delta}$ that appears in Eq.~(\ref{eq:action}), the cluster self-energy, or the cluster Green function, in a diagonal form.
Since cellular DMFT applies open boundary conditions on the cluster, the cluster momentum ${K}$ is not in general a good quantum number.
However, in a $2\times2$ plaquette, one still has C4$_v$ symmetry and the one-dimensional irreducible representations of the subgroup C2$_v$, equivalently noted by ${K}$, can be used.
Then the cluster quantities take a diagonal form with indices
$(0,0)$, corresponding to the representation A$_1$, $(\pi,\pi)$ corresponding to the representation A$_2$ and $(\pi,0)$, $(0,\pi)$ corresponding to the representations B$_1$ or B$_2$.
For example, the bath hybridization function in cluster momentum basis reads:
\begin{equation}
\hat{\Delta} = \left( \begin{array}{cccc}
                   \Delta_{(0,0)} & 0 & 0 & 0 \\
                   0 & \Delta_{(\pi,0)} & 0 & 0 \\
                   0 & 0 & \Delta_{(0,\pi)} & 0 \\
                   0 & 0 & 0 & \Delta_{(\pi,\pi)}
                  \end{array} \right).
\label{eq:Delta_basis_K}
\end{equation}
The diagonal elements are the eigenvalues of the matrix containing the on site, nearest neighbor, and next-nearest neighbor cluster quantities in real space.
Physically they can be thought as an average of the lattice quantity over a coarse grained approximation of the Brillouin zone.
In analogy with multiband models, we refer to these cluster momenta ${K}$ as ``orbitals''.
When exploiting this interpretation, one should keep in mind, however, that in the cellular DMFT scheme the cluster momenta ${K}$ are coupled through both the on-site interaction $U$ and the hopping matrix $t$, which in cluster momentum basis takes the form:
\begin{equation}
\hat{t}(\tilde{k}) = \left( \begin{array}{cccc}
                   t_{11} & t_{12} & t_{13} & 0 \\
                   t_{12}^* & t_{22} & 0 & t_{13} \\
                   t_{13}^* & 0 & t_{33} & t_{12} \\
                   0 & t_{13}^* & t_{12}^* & t_{44}
                  \end{array} \right).
\label{eq:hopping_basis_K}
\end{equation}
where $t_{11}=-t\left(2+ \cos(2k_x)+\cos(2k_y)\right)$, $t_{22}=t(\cos(2k_x)-\cos(2k_y))$, $t_{33}=-t(\cos(2k_x)-\cos(2k_y))$, $t_{44}=t(2+ \cos(2k_x)+\cos(2k_y))$, $t_{12}=it\sin(2k_x)$,  $t_{13}=it\sin(2k_y)$,  and $k_x, k_y \in [-\pi/2,\pi/2]$. 
\begin{figure}[!ht]
\centering{\includegraphics[width=0.8\linewidth,clip=]{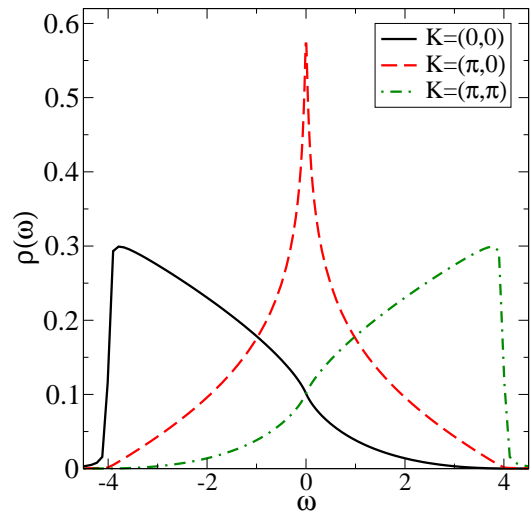}}
\caption{Non interacting local density of states $\rho_{K}(\omega)=-1/\pi ImG_{K}(\omega)$ of the orbitals $K=(0,0), (\pi,0), (\pi,\pi)$ (solid, dashed and dot-dashed lines respectively).
}
\label{fig1}
\end{figure}
For reference, the non-interacting local density of states on the orbitals obtained from
\begin{equation}
\hat{G}(\omega) = \int d\tilde{k} \left[ \omega +\mu -\hat{t}(\tilde{k}) \right]^{-1}
\label{eq:GFnonint}
\end{equation}
is shown in Fig.~\ref{fig1}.

CTQMC is an imaginary time technique, thus for the interacting case the local density of states is not directly accessible and has to be computed by analytical continuation schemes that rely on additional approximations.
Nevertheless, the characterization of the phase diagram of the normal phase of the Hubbard model -- which is the topic of this study -- can unequivocally be determined from observables directly measured in the CTQMC method, like the particle density $n$, the cluster Green's function and self-energy in Matsubara frequency.
Here we restrict our study to that kind of quantities.
From a methodological viewpoint, this is a necessary step before using other methods to extract additional information.


\section{Phase diagram}
\label{Sec:PhaseDiagram}

In this section we describe the normal-state phase diagram of the two-dimensional Hubbard model determined in detail by cellular DMFT calculations\cite{sht}.

The parameter space is three dimensional and consists of all possible values of the interaction strength $U$, the temperature $T$ and the chemical potential $\mu$ (or, equivalently, the doping $\delta$ defined as $\delta=1-n$).
To identify the different phases and locate their boundaries or crossover lines, one must scan the phase diagram with an adapted fine grid.
Despite the algorithmic breakthroughs described in the previous section, that have become available only in the last few years, it took several million of CPU hours to be able to determine the whole normal-state phase diagram of the two-dimensional Hubbard model.

\begin{figure*}[!ht]
\centering{\includegraphics[width=0.9\linewidth,clip=]{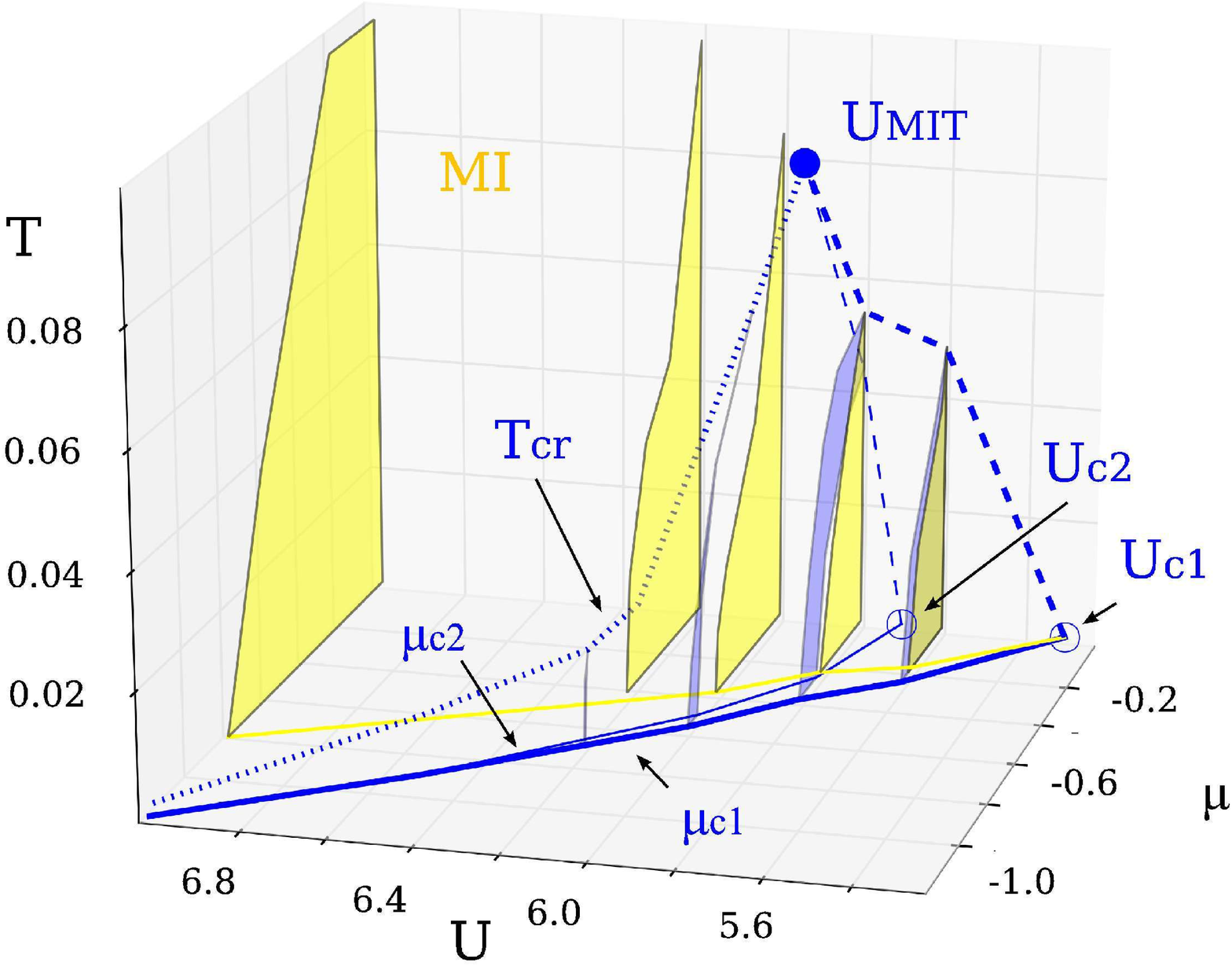}}
\caption{Chemical potential $\mu$, interaction $U$, temperature $T$ phase diagram of the two-dimensional Hubbard model obtained by cellular DMFT. Because of particle-hole symmetry it is symmetric with respect to the $\mu=0$ plane. Cross-sections at constant $U$ are shown. Dark-gray (blue) shaded regions represent the coexistence of two phases. Light-gray (yellow) areas denote the onset of the Mott insulator state (MI), characterized by a plateau in the occupation at $n=1$. When these two regions overlap, a metal-insulator transition takes place (different shade of gray). Otherwise, the coexistence regions occur between two different metals. Projections on $T=0$, and $\mu=0$ planes are also shown (full lines and dashed lines respectively). Open dots mark the extrapolated $T=0$ values of $U_{c1}$ and $U_{c2}$.
A critical line $T_{\rm cr}$ (dotted line) originates at the half-filled Mott critical endpoint $U_{\rm MIT}$ (full dot) and moves to progressively low temperatures and high doping as $U$ increases.
}
\label{fig2}
\end{figure*}
\begin{figure*}[!ht]
\centering
\subfigure[]{\label{fig3a}\includegraphics[width=0.24\linewidth]{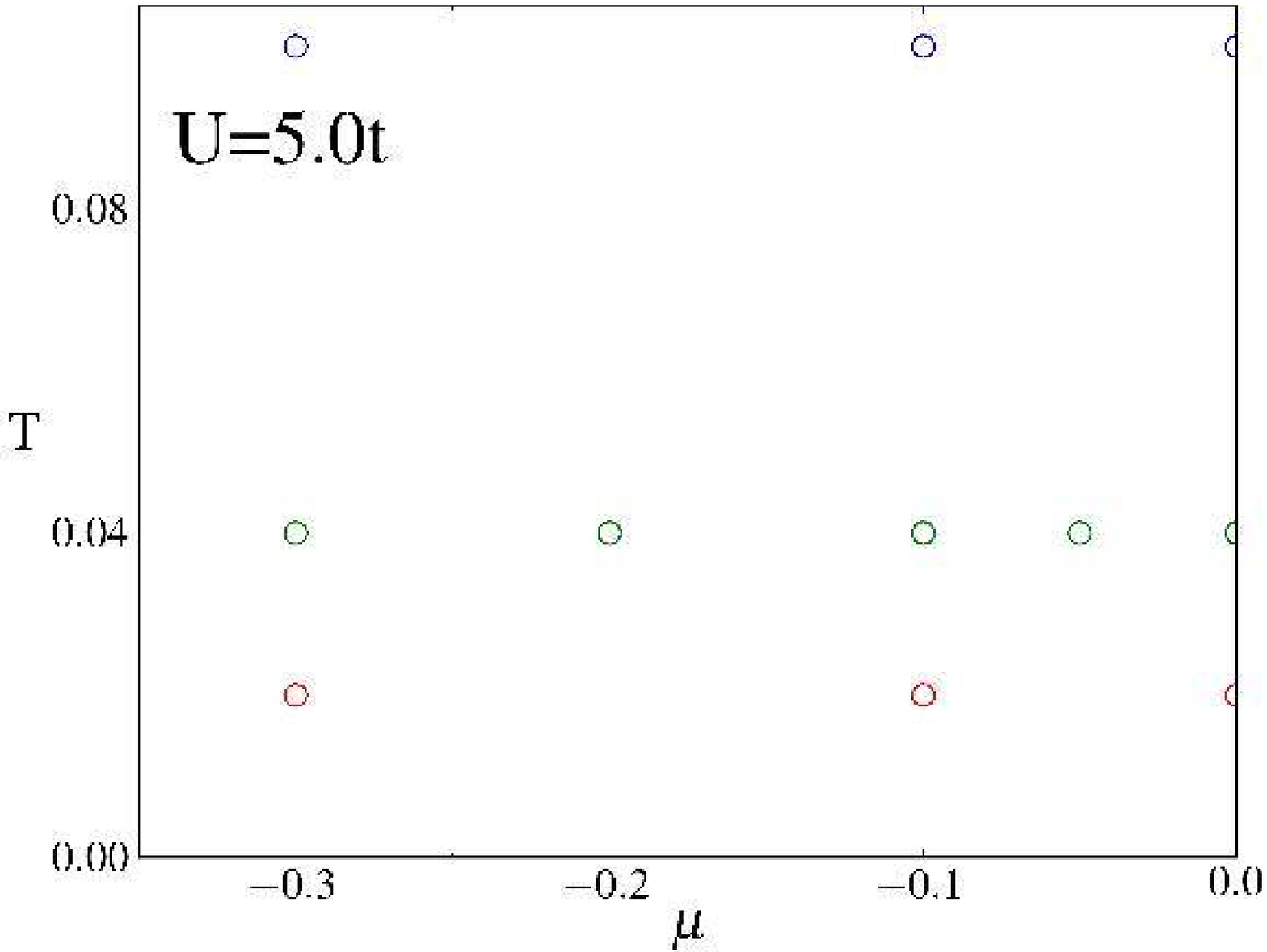}}
\subfigure[]{\label{fig3b}\includegraphics[width=0.24\linewidth]{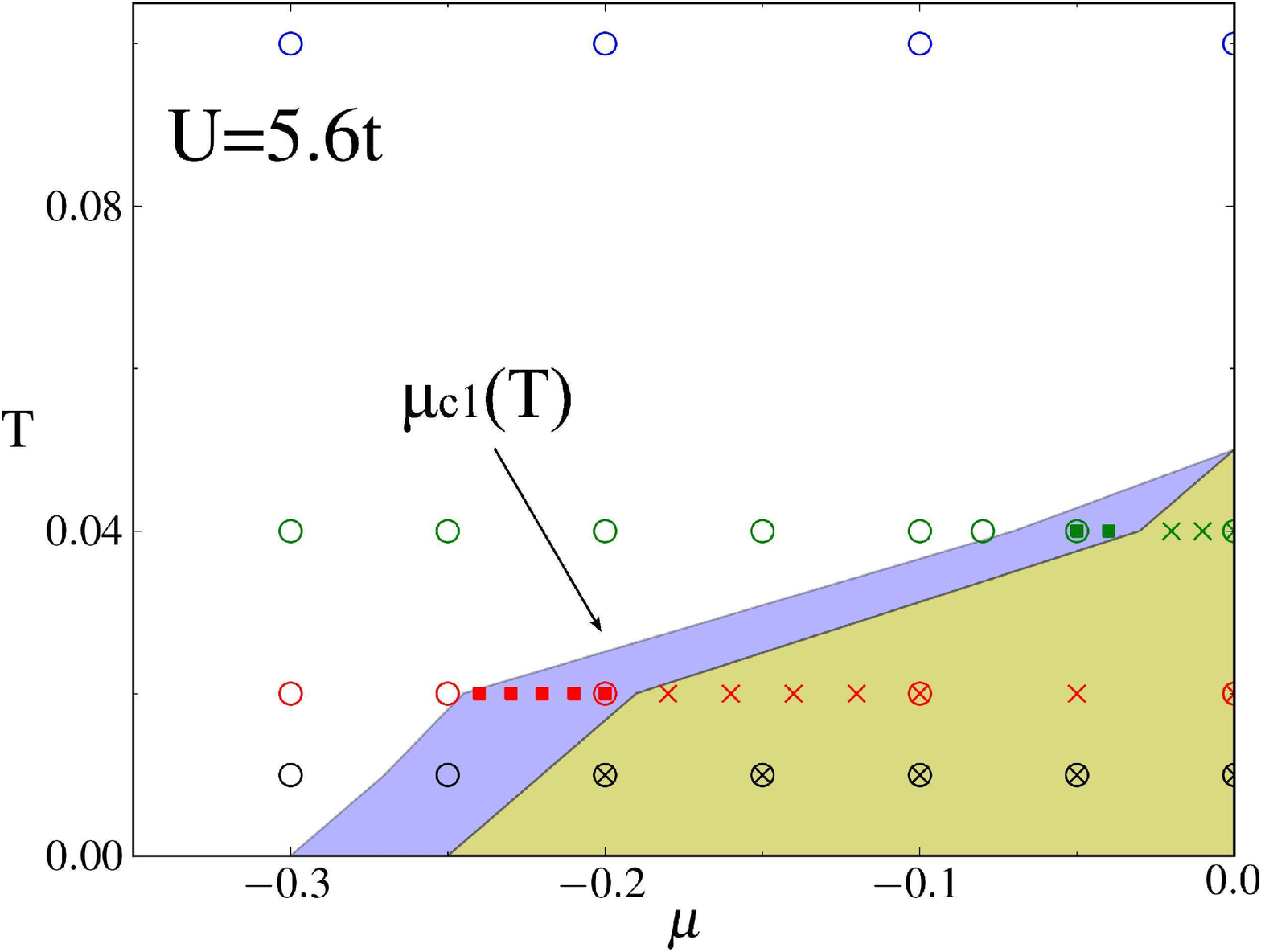}}
\subfigure[]{\label{fig3c}\includegraphics[width=0.24\linewidth]{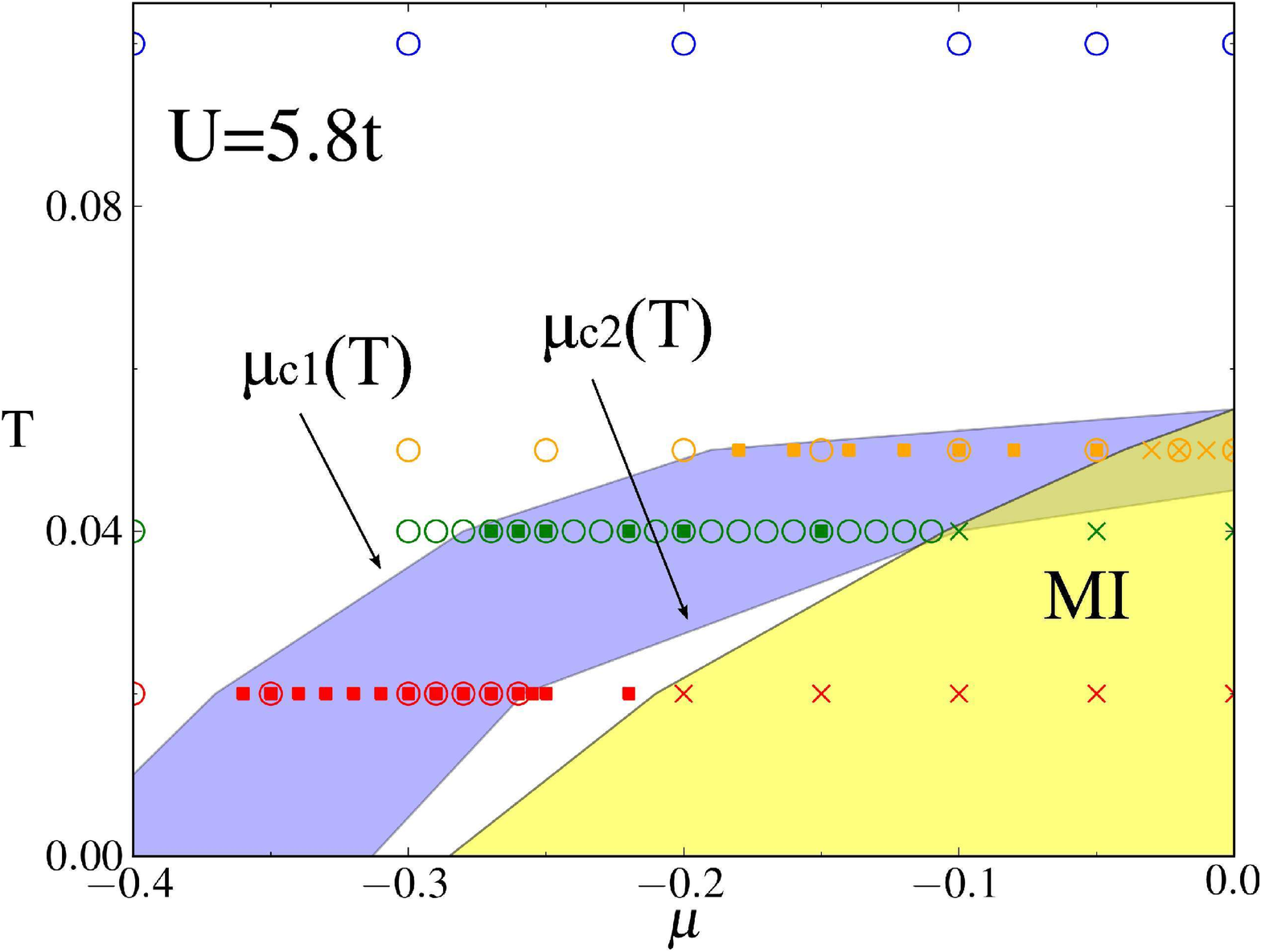}}
\subfigure[]{\label{fig3d}\includegraphics[width=0.24\linewidth]{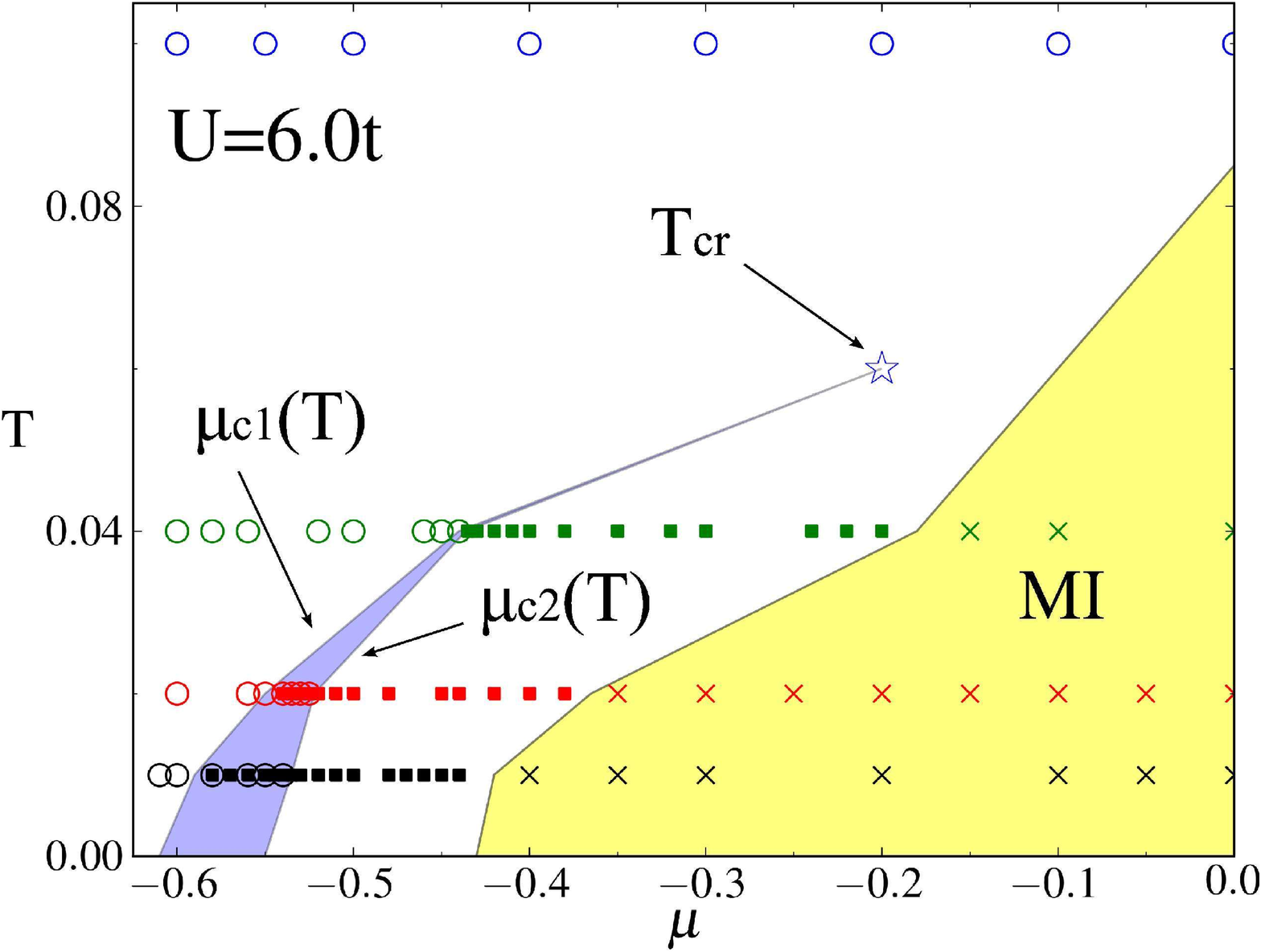}}

\subfigure[]{\label{fig3e}\includegraphics[width=0.24\linewidth]{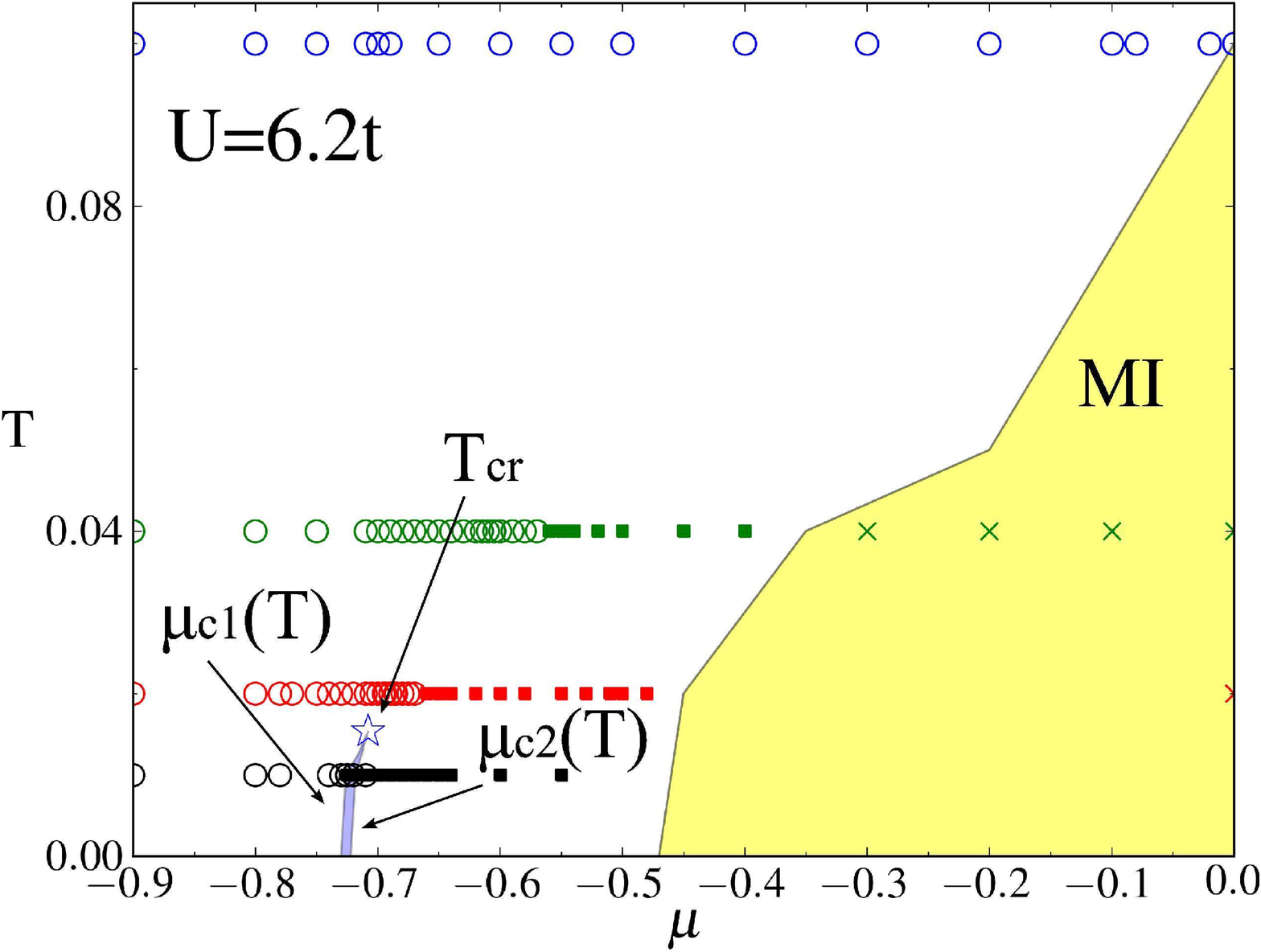}}
\subfigure[]{\label{fig3f}\includegraphics[width=0.24\linewidth]{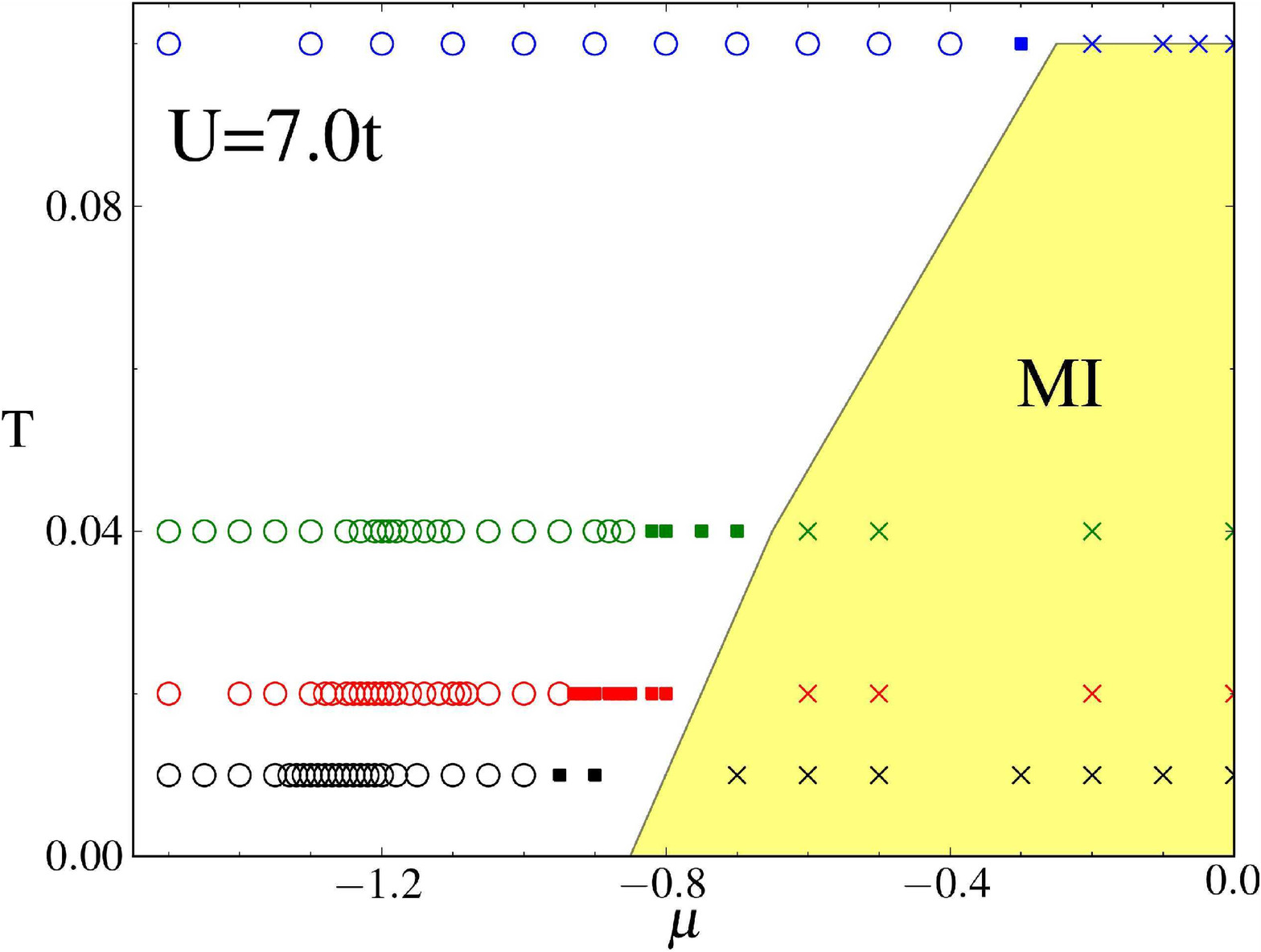}}
\subfigure[]{\label{fig3g}\includegraphics[width=0.24\linewidth]{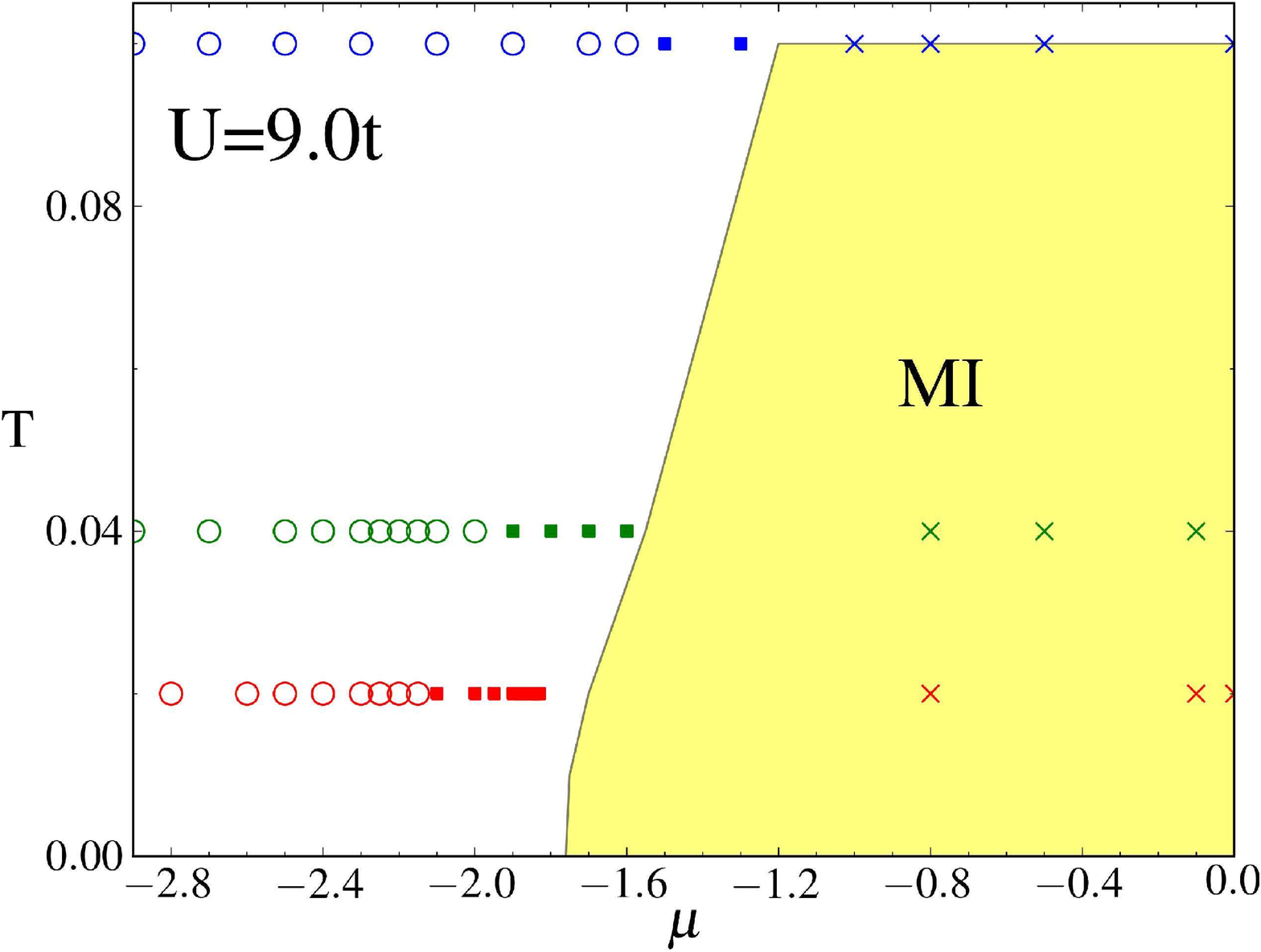}}
\subfigure[]{\label{fig3h}\includegraphics[width=0.24\linewidth]{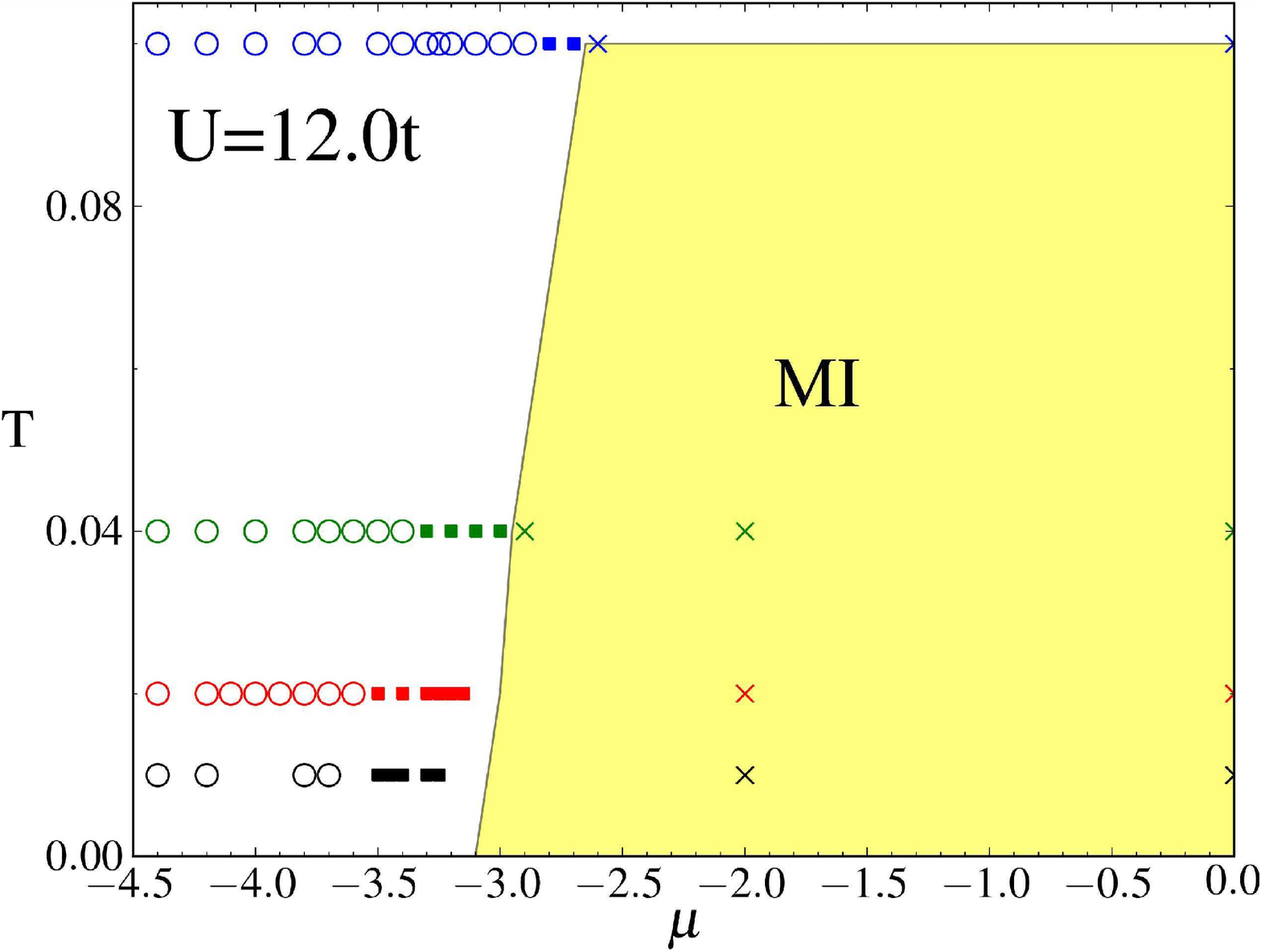}}
\caption{Phase diagram in the temperature $T$ versus chemical potential $\mu$ plane for different values of the interaction strength $U$. Each panel corresponds to a cross-section at constant $U$ of the previous figure. As before, light gray (yellow) area represents the Mott insulating phase (MI). Dark grey (blue) area represents the coexistence of two phases and is bounded by the spinodal lines $\mu_{c1}(T)$ and $\mu_{c2}(T)$. A star symbol marks the end of the coexistence region at finite temperature; by following this point as a function of $U$ gives rise to the dotted line $T_{\rm cr}$ of Fig.~\ref{fig2}. Other symbols denote some of the points actually computed in our study. Three phases can be distinguished: overdoped phase (open circles), Mott insulator (crosses), underdoped phase (filled squares). The former is metallic. In the latter some regions of the Brillouin zone are gapped and others gapless, and the system is compressible.
}
\label{fig3}
\end{figure*}

Specifically, we carried out scans at constant values of $U$ across the Mott transition, varying the chemical potential for several temperatures.
Because the system is particle-hole symmetric ($t'$ is set to zero), we limited our study to hole doping only ($\mu<0$, $\delta<1$).
For each value of $U$, we performed calculations in a wide doping range, between $0$ and roughly $25\%$.
We found that scanning doping in intervals sometimes smaller as $0.002$ was necessary to identify the sequence of phases that occur upon doping the Mott insulator.
As far as temperature is concerned, we mostly focused on the low temperature regime in the decade $1/100<T/t<1/10$.

We summarize our results in the temperature $T$ versus interaction strength $U$ and chemical potential $\mu$ phase diagram shown in Fig.~\ref{fig2}.
The scans at constant $U$ that we use to explore the three dimensional parameter space are visible as cross sections in the phase diagram and correspond to the distinct $T-\mu$ planes presented in Fig~\ref{fig3}.

The figures show two colored (shaded) regions in cross sections at constant $U$ in $(U,T,\mu)$ space.
The first region (light gray/yellow) corresponds to the onset of the Mott insulating phase, characterized by a plateau in the occupation at $n=1$.
Just outside this region there is a metallic state.
The second region (dark gray/blue) is the portion of parameter space where two different phases coexist.
By interpolation between the distinct planes at constant $U$, one can therefore obtain the volume of both the Mott phase and the phase coexistence region.

It is important to understand that what we mean by coexistence region is a region where, in addition to the thermodynamically stable phase, a metastable phase can exist. Since we do not have access to the free energy, we cannot determine precisely the first order boundary, so instead we draw the region where metastable states exist. In practice, if we change chemical potential slowly from small to large values or vice-versa, we can end up in different phases, one of which is thermodynamically unstable.

Going back to our phase diagram, let us first concentrate on the Mott phase.
Two basic features emerge.
First, as expected on physical grounds, a threshold value for the correlation strength, called $U_{c1}$, is a prerequisite for the emergence of Mott insulating state.
We estimate this point at $(U,T,\mu)\approx(5.3t,0,0)$.
Second, the projections on $T=0$ plane (marked as light gray/yellow line on the 3D phase diagram) reveal the characteristic $V$ shape onset of the Mott insulating state, where the tip of the $V$ shape is $U_{c1}$ (only half of the $V$ shape is shown in Fig.~\ref{fig2} because of particle-hole symmetry).

Our most surprising results come, however, from the full mapping of the region (dark gray/blue) where two phases coexist in the $(U,\mu,T)$ space.
We begin with a description of the peculiar shape of this region.
Let first consider the $\mu=0$ plane, where the model is half filled.
In that plane, the coexistence region has a triangular shape that bends as $U$ is increased. The boundaries are the spinodal lines $U_{c1}(T)$ (thick dashed blue line), where the metastable insulating phase disappears, and $U_{c2}(T)$ (thin dashed blue line), where the metastable metallic phase ceases to exists.
These spinodals end at a critical value of $U$, called $U_{\rm MIT}$, with coordinates $(U,T,\mu)\approx(5.95,0.08,0)$.
In the $\mu=0$ plane, the region of coexistence marks a portion of parameter space where the Mott insulator coexists with a metal.

Surprisingly, this is in general not so in the 3D phase diagram where the coexistence region is not fully included in the region where the Mott phase exists.
To show this, let us consider scans at constant $U$.
The region where two phases coexist naturally extends from the $\mu=0$ plane to finite values of the chemical potential $\mu$ and is enclosed by the spinodal surfaces $\mu_{c1}(U,T)$ and $\mu_{c2}(U,T)$ that end at a critical line $T_{\rm cr}$ (dotted blue line in Fig.~\ref{fig2}). 
That line is the continuation of the critical point $(U,T,\mu)\approx(5.95,0.08,0)$ from half filling to finite doping.
The dark grey (blue) shaded areas in the phase diagrams correspond to the intersection between the volume delimited by these surfaces and the cross sections at constant $U$.

Let us look in more detail at the coexistence regions. 
In Fig.~\ref{fig3} we show constant $U$ cuts of Fig.~\ref{fig2}.  
For values of the interaction strength $U$ larger than $U_{\rm MIT}$, the region of coexisting phases displays again a triangular shape bending towards the Mott insulator (i.e. towards low doping). As we discuss in the next section, two metallic phases coexist in that region.
The coexistence begins at the temperature $T_{\rm cr}$, marked by a star symbol in Fig.~\ref{fig3}, and is delimited by the spinodals $\mu_{c1}(T)$ and $\mu_{c2}(T)$.
This is apparent from the scans at $U/t=6.0,6.2$. For larger $U$, the coexistence region narrows and drops to temperatures lower than our current numerical capabilities.
On the other hand, as the interaction strength $U$ decreases towards $U_{\rm MIT}$, the triangular region grows and approaches the $\mu=0$ plane and its summit at the top reaches that plane at $U_{\rm MIT}$.
As $U$ is further decreased in the range $U_{c2}(T=0) < U < U_{\rm MIT}$, the coexistence regions show a rainbow-like shape (cf scan at $U=5.8t$) with the $\mu_{c1}(T)$ line on the outer part of the arc and the $\mu_{c2}(T)$ line on the inner part of the arc.
Further lowering $U$ in the region $U_{c1}(T=0) < U < U_{c2}(T=0)$, only the spinodal $\mu_{c1}$ exists until $U_{c1}(T=0)$ where it vanishes.

Our results can be summarized as follows.
First, within the volume in $(U-\mu-T)$ parameter space enclosed by the spinodal surfaces $\mu_{c1}(T,U)$ and $\mu_{c2}(T,U)$, one can obtain two solutions of the cellular DMFT equations.
Therefore a first-order transition surface has to occur in the blue region.
The precise location of the first order surface can be computed by comparing the free energy of the two solutions, a task beyond the present work.

Second, the two solutions merge at a finite temperature $T_{\rm cr}$ where the spinodal surfaces $\mu_{c1}(T,U)$ and $\mu_{c2}(T,U)$ end.
This implies that a second-order transition line (dotted blue line in Fig.~\ref{fig2}) exists at the end of the surface of first-order transition.
Except for the critical Mott endpoint $U_{\rm MIT}$, the second-order critical line takes place {\it outside} the Mott insulating phase, so it occurs for finite values of doping.

Third, our most striking finding is the nature of the phases that are found to coexist.
The first-order transition surface originates at the half-filled Mott critical endpoint and moves progressively away from half-filling as the interaction $U$ increases. 
Therefore the first-order transition can occur between a metallic phase and a Mott insulating phase, or, unexpectedly, between two metallic phases, one of which evolves continuously from the Mott insulator.
The former case (metal to insulator transition) occurs whenever the volume of the Mott insulating phase intercepts the volume of the phase coexistence in the $(U,\mu,T)$ parameter space.
As discussed above, this occurs at the $\mu=0$ plane (i.e. at half filling), where the metallic and the insulating solutions of the cellular DMFT equations coexist.
In the $T-\mu$ planes too, the Mott insulator can coexist with a metal, as indicated in Figs.~\ref{fig2},\ref{fig3} by different shade of gray (corresponding to the light gray/dark gray -- yellow/blue overlap) for $U/t=5.6,5.8$.
Even though in some region of the phase diagram the first-order transition occurs between a metal and a Mott insulator, this is generally not the case.
For sufficiently large $U$ ($U> U_{\rm MIT}$), there is a first-order transition between two metals, as can be seen in Fig.~\ref{fig2} where for $U> U_{\rm MIT}$ there is no intercept between the coexistence region (dark gray/blue) and Mott region (light gray/yellow).
In all cases shown in the phase diagram, however, the $\mu_{c2}(U,T)$ spinodal surface denotes the vanishing of a metallic solution, while the $\mu_{c1}(U,T)$ spinodal surface does not necessarily coincide with the end of Mott insulating phase.

In the following, we refer to the metallic phases separated by the first-order transition as underdoped (UD) and overdoped (OD) phase.
Here these terms indicate whether the level of doping of the phase is below or above the level of doping at which the first-order transition occurs. The underdoped phase is closest to the Mott insulator.
The connection of these names to the maximum of the superconducting dome will be discussed in Sec.~\ref{Sec:Discussion}.

\section{First-order transition at finite doping}
\label{Sec:Transition}

\begin{figure*}[!ht]
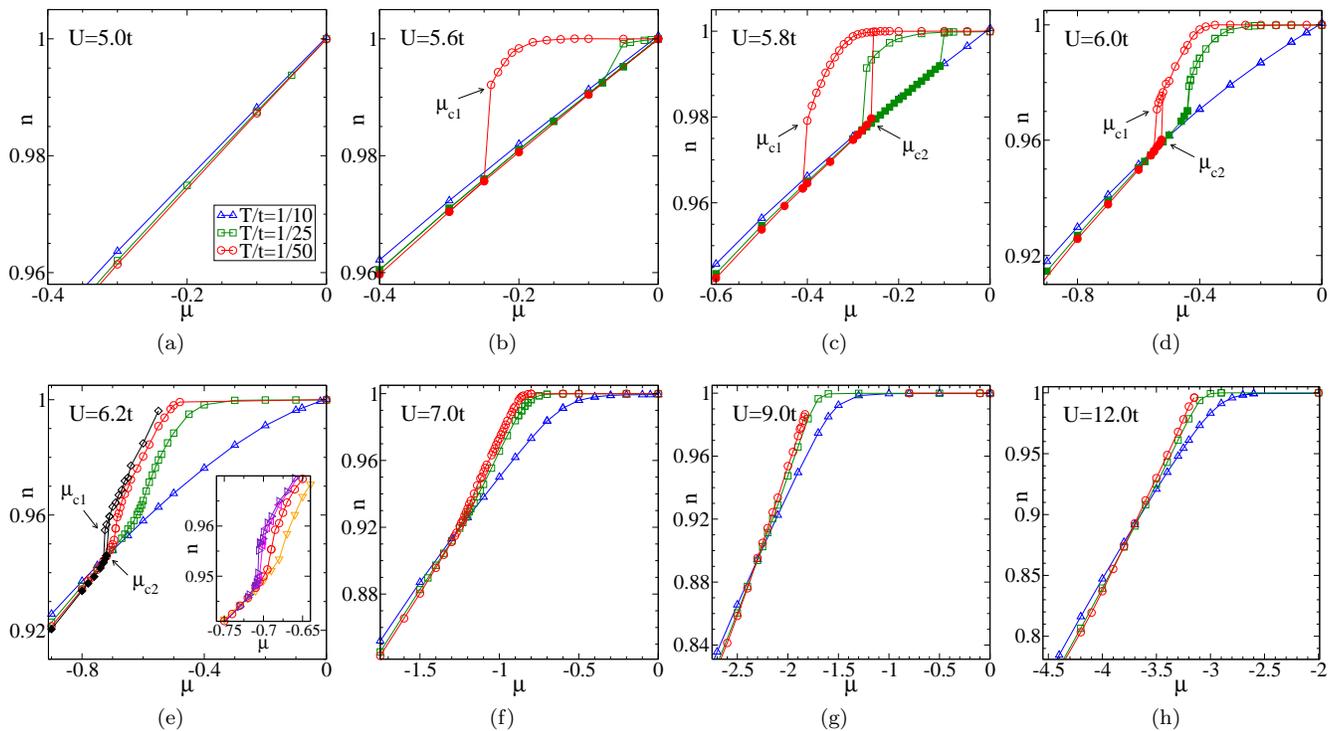

\centering
\subfigure[]{\label{fig4a}\includegraphics[width=0.24\linewidth]{fig4a.eps}}
\subfigure[]{\label{fig4b}\includegraphics[width=0.24\linewidth]{fig4b.eps}}
\subfigure[]{\label{fig4c}\includegraphics[width=0.24\linewidth]{fig4c.eps}}
\subfigure[]{\label{fig4d}\includegraphics[width=0.24\linewidth]{fig4d.eps}}

\subfigure[]{\label{fig4e}\includegraphics[width=0.24\linewidth]{fig4e.eps}}
\subfigure[]{\label{fig4f}\includegraphics[width=0.24\linewidth]{fig4f.eps}}
\subfigure[]{\label{fig4g}\includegraphics[width=0.24\linewidth]{fig4g.eps}}
\subfigure[]{\label{fig4h}\includegraphics[width=0.24\linewidth]{fig4h.eps}}
\caption{Occupation $n$ versus $\mu$ for several values of interaction strength $U$. The data shown are for temperatures $T/t=1/10$ (blue triangles), $1/25$ (green squares), $1/50$ (red circles), $1/100$ (black diamonds). When two solutions are found to coexist, the solutions obtained following the metallic and the insulating solution are indicated as full and open symbols respectively.  $\mu_{c1}$ marks the vanishing of the underdoped phase. $\mu_{c2}$ signals the disappearance of the overdoped phase. The arrows indicate the values of $\mu_{c1}$ and $\mu_{c2}$ at the lowest temperature shown in the panel.
The plateau in the occupation at $n=1$ signals the onset of the incompressible Mott state. Note that $\mu_{c1}$ in general occurs at finite values of doping. The inset in the panel for $U=6.2t$ shows the $n(\mu)$ curves for temperatures $T/t=1/40$ (orange down triangles), $1/50$ (red circles), $1/60$ (magenta left triangles), $1/64$ (violet right triangles), which are above the second order critical temperature $T_{\rm cr}\sim 1/65t$ at which the two phases merge. Note the sigmoidal shape of the $n(\mu)$ curves as $T_{\rm cr}$ is approached from above.
}
\label{fig4}
\end{figure*}

The most important results in the normal-state phase diagram surveyed so far is the first-order transition surface in the 3D parameter space $(U,T,\mu)$ with the different phases it separates.
The focus of this section is a systematic analysis of this first-order transition.
We begin with the numerical results that demonstrate the existence of such a transition, and then turn to the thermodynamic behavior of several observables near its boundary.

Fig.~\ref{fig4} demonstrates the existence of a finite $T$ first-order transition.
We plot the occupation $n$ as a function of the chemical potential $\mu$ for a wide range of temperature and eight values of $U$.
It is important to stress that each curve $n(\mu)$ is obtained by keeping all model parameters fixed except $\mu$.
A plateau in the curves at $n(\mu)=1$ appears above the critical coupling $U_{c1}(T)$ and signals the incompressible Mott insulating phase.
The marked temperature dependence of the $n(\mu)$ curves in the UD phase, as opposed to the OD phase, demonstrates that the effects of strong correlations manifest themselves at low energy in the UD phase.

For $U/t=5.6,5.8,6.0,6.2$ and low temperatures, the occupation clearly shows hysteretic behavior as a function of $\mu$, an unequivocal fingerprint of the first-order nature of the transition. 
To find the hysteresis loop, we use an iterative procedure on a fine grid of $\mu$ points, where a converged solution is used as seed for the next value of $\mu$.
We obtain the {\it lower} branch of the hysteresis cycle by increasing $\mu$ starting from large enough doping.
This branch corresponds to the metallic branch of the overdoped phase and is indicated by full symbols in Fig.~\ref{fig4}.
For $U/t=5.6$, which lies in the region $U_{c1}(T=0)<U<U_{c2}(T=0)$, this branch continues up to half filling, at $\mu=0$.
In contrast, for $U$ larger than $U_{c2}(T=0)$, the metallic branch of the OD phase shows an upward jump at finite doping.
The OD metallic branch endpoint defines $\mu_{c2}$.
On the other hand, to compute the {\it upper} branch of the hysteresis loop, we decrease $\mu$ starting from the Mott insulating solution at half filling.
This branch is indicated by open symbols in Fig.~\ref{fig4}.
As $\mu$ varies in the Mott plateau, this branch remains constant at $n=1$, then it evolves continuously into a compressible metal (i.e. the underdoped phase), and subsequently undergoes a downward jump at finite doping.
The endpoint of the upper branch defines $\mu_{c1}$.

The hysteresis region delimits a region of space parameters $(U-T-\mu)$ where two solutions of the cellular DMFT equations coexist.
The crossing of the free energy of the two solutions in this volume bounded by $\mu_{c1}(T,U)$ and $\mu_{c2}(T,U)$ defines the first-order transition surface.
Determining this surface and finding out which is the most stable phase in the coexistence region necessitates to compute the free energy. This task requires a large investment in computing time and is beyond the scope of the present work.

The continuous evolution from the Mott plateau to the compressible UD phase seems to occur towards a discontinuous change in the compressibility (first derivative of $n(\mu)$) at $T=0$ which would correspond to a second order transition, or quantum critical point. This is apparent especially at large $U$. At finite temperature, only higher order derivatives are perhaps discontinuous.

The $\mu_{c2}(T,U)$ surface always coincide with the vanishing of the metallic state (namely, the OD phase), as revealed by the jump of the lower branch at finite doping.
In sharp contrast, the $\mu_{c1}(T,U)$ surface does not in general mark the vanishing of the Mott insulating state: as shown in Fig.~\ref{fig4}, the sudden jump at $\mu_{c1}$ occurs at finite doping!
These two observations lead to the conclusion that the first-order transition can actually occur between two types of metallic phases. Since these phases have the same symmetry, the first-order surface can end at a critical line at finite temperature $T_{\rm cr}$ (dotted blue line in the 3D phase diagram of Fig.~\ref{fig2}).
As $T_{\rm cr}$ is approached from below, the hysteresis loop narrows and vanishes at $T_{\rm cr}$.
On the other hand, as $T_{\rm cr}$ is approached from above, the curves show a sigmoidal profile, a clear signature of proximity to a second-order critical point.
The inset of Fig.~\ref{fig4e} reveals this remarkable feature.

Hysteresis in the occupation curves $n(\mu)$, found for $U/t=5.6,5.8,6.0,6.2$ and low temperatures, provides a direct demonstration of the first-order nature of the transition.  
The curves at $U/t=7.0, 9.0, 12.0$ do not show hysteresis down to $T/t=1/100$, but several signatures indicate that the system is getting close to the transition and hence suggest that the hysteresis loop falls below the lowest temperature we investigate. 
To reach this conclusion, first we notice that a close inspection of the $n(\mu)$ curves at $U/t=6.0,6.2$, which lie above the critical interaction $U_{\rm MIT}\approx5.95t$, shows that the temperature $T_{\rm cr}$ dramatically decreases as $U$ increases.
Indeed, the critical line begins at the Mott point at $T/t\sim 0.08$ and decreases to $T/t\sim0.015$ in going from $U_{\rm MIT}\approx5.95t$ to $U=6.2t$.
Second, by examining the high-temperature precursors of the transition, we can infer about the existence of the first-order transition at large values of $U$. 
A thermodynamic indicator that signals the proximity to the second-order transition line is a peak in the compressibility $(dn/d\mu)|_T$ as a function of doping that develops at temperatures well above $T_{\rm cr}$ and reaches its maximum at $T_{\rm cr}$ (see also discussion of Fig.~\ref{fig16}). 
We found this precursor of $T_{\rm cr}$ for all values of $U>U_{\rm MIT}$ at progressively larger doping as $U$ increases. 
Another signature that can be identified as a precursor of the transition is the approximate vanishing of the expansion coefficient $(dn/dT)|_{\mu}=0$ (crossing point of the isotherms in Fig.~\ref{fig4}). This signal occurs close to the spinodal line $\mu_{c2}(T)$ and extends to temperature well above $T_{\rm cr}$. This feature occurs for all values of $U>U_{c2}$ and at progressively larger doping as $U$ increases.

Our results show these high temperature precursory signatures of the transition (other indicators will be discussed is Sec.~\ref{Sec:Discussion}) for all values of the correlation strength $U>U_{\rm MIT}$ and hence point toward a critical temperature decreasing with $U$ and occurring at progressively larger doping as $U$ increases.
We estimate that the critical doping at which $T_{\rm cr}$ occurs moves from zero doping at the Mott point $U_{\rm MIT}\approx5.95t$, to $n\approx9.95$ at $U=6.2t$, and to $\approx0.88$ at $U=12.0t$. 
Some of the thermodynamic effects close to $T_{\rm cr}$ will be analyzed in more detail in the next section.
Here we note that the physics arising from the second-order critical line strongly modifies not only the thermodynamic response, but also the dynamics: as we shall see in Sec.~\ref{Sec:Phases}, there is a funnel-shaped region above the critical line where scattering is anomalously large.

Such precursory signals strongly indicate the occurrence of the first-order transition surface ending at a finite temperature second-order transition line for $U>U_{\rm MIT}$ and thus also for $U>6.2t$ for temperature lower than the range we explored. 
In addition, these precursory signals take place not only for all values of $U>U_{\rm MIT}$, but also without apparent qualitative change. 
This suggests to disfavor other interpretations where a qualitative change in these high temperature signatures is expected, like the possibility that the critical line ending our first order transition becomes a quantum critical line or point~\cite{imadaCR1,imadaCR2}, or a tricritical point at some $U$. 

Additional support for the approach of a finite $T$ second-order transition line follows from the critical slowing down observed in our CTQMC calculations close to that line.
Critical slowing down is a widespread and standard indicator~\cite{warnings} that the system is near a critical threshold. We find enhancement of Monte Carlo fluctuations and also enhancement of the number of the cellular DMFT iterations to attain self-consistency as the model parameters approach $T_{\rm cr}$.
To obtain reliable results, a large number of Monte Carlo samplings and of the order of hundreds of iterations are used in our numerical simulations.
This effect is well documented in single-site DMFT studies of the Mott transition~\cite{oudo,sahana}.
In cellular DMFT investigations, it is reported in Ref.~\onlinecite{hauleDOPING} at $U=12.0t$ around $n\sim0.88$; here we detect this phenomenon down to $U_{\rm MIT}$ at progressively smaller doping, hence linking the critical slowing down to Mott physics.

\section{Thermodynamics of the transition}
\label{Sec:Thermodynamics}

Having established the first-order character of the transition, we now focus on the thermodynamic properties of the phases that this transition separates.
We discuss four basic features.

First we prove that these phases are thermodynamically stable.
We start from the fundamental thermodynamic relation
\begin{equation}
dE(S,n,U)=TdS +\mu dn +DdU,
\label{eq:dE}
\end{equation}
where it is understood that energy $E$, entropy $S$, number of particle $n$ and double occupancy $D$ are normalized per lattice site.
The model parameters are $T,\mu,U$, so it is natural to use the grand canonical potential $\Omega$.
The appropriate Legendre transform leads to
\begin{equation}
d\Omega = d(E-TS-\mu n) = -SdT -nd\mu +DdU.
\label{eq:defdO}
\end{equation}
From the expression for $d\Omega$ in Eq.~(\ref{eq:defdO}) and $dU=0$ from now on, we have
\begin{equation}
\left(\frac{\partial\Omega}{\partial T}\right)_{\mu} = -S \; ; \qquad \left(\frac{\partial\Omega}{\partial \mu}\right)_{T} = -n.
\end{equation}
The concavity of the entropy implies that the grand canonical potential is minimum at equilibrium. Thus:
\begin{equation}
\begin{split}
d^2\Omega &= \left(\frac{\partial^2 \Omega}{\partial T^2}\right)_{\mu} (dT)^2
            +2\left(\frac{\partial^2 \Omega}{\partial T \partial\mu}\right) dT d\mu
            +\left(\frac{\partial^2 \Omega}{\partial \mu^2}\right)_{T} (d\mu)^2 \\
          &= \begin{pmatrix} dT & d\mu \end{pmatrix}
             \begin{pmatrix} \left(\frac{\partial^2 \Omega}{\partial T^2}\right)_{\mu} & \left(\frac{\partial^2 \Omega}{\partial T \partial\mu}\right) \\  \left(\frac{\partial^2 \Omega}{\partial T \partial\mu}\right) & \left(\frac{\partial^2 \Omega}{\partial \mu^2}\right)_{T} \end{pmatrix}
             \begin{pmatrix} dT & d\mu \end{pmatrix} < 0.
\end{split}
\end{equation}
If we consider a constant $T$ plane the last relation is obeyed if
\begin{equation}
\left(\frac{\partial^2 \Omega}{\partial\mu^2}\right)_{T} = -\left(\frac{\partial n}{\partial\mu}\right)_{T} < 0
\end{equation}
i.e. if the compressibility is positive.
In the general case, we need to add the requirement that the eigenvalues of the above matrix be negative, i.e that the determinant be positive:
\begin{equation}
\left(\frac{\partial^2 \Omega}{\partial T^2}\right)_{\mu}  \left(\frac{\partial^2 \Omega}{\partial \mu^2}\right)_{T} -\left(\frac{\partial^2 \Omega}{\partial T \partial\mu} \right)^2 > 0.
\label{eq:stability}
\end{equation}
Notice that the specific heat
\begin{equation}
-T \left(\frac{\partial^2 \Omega}{\partial T^2}\right)_{\mu} = T \left(\frac{\partial S}{\partial T}\right)_{\mu}
\end{equation}
must be positive as a consequence of the above two inequalities.
Given
\begin{equation}
\left(\frac{\partial^2 \Omega}{\partial T \partial\mu} \right) =-\left(\frac{\partial n}{\partial T} \right)_{\mu}
\end{equation}
and the expression for $dE$, Eq.~\ref{eq:dE}, we can extract the specific heat from our data since it has the equivalent expression:
\begin{equation}
T \left(\frac{\partial S}{\partial T}\right)_{\mu}
= \left(\frac{\partial E}{\partial T}\right)_{\mu} - \mu \left(\frac{\partial n}{\partial T} \right)_{\mu}.
\end{equation}
\begin{table}
\begin{tabular}{|c|c|c|c|c|}
  \hline
  $n$ & $\left(\frac{\partial n}{\partial \mu} \right)_{T}$ & $\left(\frac{\partial n}{\partial T} \right)_{\mu}$ & $\left(\frac{\partial E}{\partial T}\right)_{\mu}$ & det \\
  \hline
  0.982 & 0.201 & -0.632 & -1.237 & 4.420 \\
  0.971 & 0.272 & -0.943 &  -1.899 & 6.944 \\
  0.941 & 1.146 & 0.116 & 0.585 & 3.170 \\
  0.921 & 0.128 & 0.055 & 0.648 & 4.725 \\
  \hline
\end{tabular}
\label{tab:tab1}
\caption{Particle density $n$, compressibility $dn/d\mu|_{T}$, expansion coefficient $dn/dT|_{\mu}$, derivative of total energy $dE/dT|_{\mu}$ and the determinant Eq.~\ref{eq:stability}. The first (last) two lines correspond to values of $n$ in the UD (OD) phase. Other data parameters are: $U=6.2t>U_{\rm MIT}$, and the derivatives are taken at $T/t=1/70$ which lies below $T_{\rm cr}$. The determinant is positive, implying the thermodynamic stability of both the UD and OD phases.
}
\end{table}

Clearly Fig.~\ref{fig4} shows that $dn/d\mu >0$. 
As an example, we display these quantities for a few points in Table I to demonstrate that the stability condition is satisfied.
The data are for $U=6.2t>U_{\rm MIT}$ and $T/t=1/70$ which lies below $T_{\rm cr}\approx 1/65$. The determinant Eq.~\ref{eq:stability} and $dn/d\mu$ are positive for all values displayed.

Second, the surprising result that emerges from our calculations is the possibility of a first-order transition between two paramagnetic metallic phases.
The filling $n$ is one direct way to discriminate the two phases that coexist below the second-order transition line: the OD phase at lower filling and the UD phase at higher filling.

Third, the Clausius-Clapeyron relations reveal that the UD phase has smaller entropy and smaller double occupancy than the OD phase.
The proof works as follows. Along the first order transition boundary,
the grand potential $\Omega$ is identical in the two phases.
This means that if we move along the first-order transition line on either side, the change in $\Omega$ is the same in both phases: $d\Omega_{\rm UD}=d\Omega_{\rm OD}$.
If we consider a constant $U$ plane ($dU=0$) and indicate by subscript c quantities measured along the first order transition line, it immediately follows that
\begin{equation}
\frac{dT_c}{d\mu_c} = \frac{n_{\rm UD}-n_{\rm OD}}{S_{\rm OD} -S_{\rm UD}}.
\end{equation}
Our calculation shows that $T_c$ increases as $\mu_c$ increases (i.e. the first-order line bends toward the Mott insulator).
This implies that the UD phase has a lower entropy than the OD phase.
In an analogous way, by taking a constant $T$ plane, one obtains
\begin{equation}
\frac{dU_c}{d\mu_c} = \frac{n_{\rm UD}-n_{\rm OD}}{D_{\rm UD} -D_{\rm OD}}.
\end{equation}
Our calculations show that $\mu_c$ decreases as $U_c$ increases. Hence, the UD phase has lower double occupancy than the OD phase.
This is as expected and suggests again that in the UD phase the correlations are stronger.

Finally, it is important to stress that the entropy exhibits a maximum as a function of doping or chemical potential in close proximity to the spinodal surface $\mu_{c2}(U,T)$ and its high temperature crossover,.
This can be seen as follows.
Above the critical coupling $U_{c2}(T)$, and for the temperature range explored in our work, the isotherms in the $n-\mu$ plane in Fig.~\ref{fig4} approximately cross at a finite value of doping.
That doping where the isotherms cross increases with $U$.
The crossing of the isotherms translates into a vanishing expansion coefficient $(dn/dT)_{\mu}=0$ and thus into an extremum in the entropy $(dS/d\mu)_T=0$ since from the grand potential Eq.~\ref{eq:defdO} we have the Maxwell relation
\begin{equation}
(\partial S/\partial \mu)_{T,U} = (\partial n/\partial T)_{\mu,U}.
\end{equation}
As long as the compressibility $(\partial \mu/\partial n)_{T,U}$ is not singular, this also implies $(\partial S/\partial n)_{T,U}=0$ because from $S=S(T,\mu,U)$ we can write
\begin{equation}
dS = (\partial S/\partial T)_{\mu,U} dT +(\partial S/\partial \mu)_{T,U} d\mu +(\partial S/\partial U)_{T,\mu} dU
\end{equation}
that in turn leads to
\begin{equation}
(\partial S/\partial n)_{T,U} = (\partial S/\partial \mu)_{T,U} (\partial \mu/\partial n)_{T,U}.
\end{equation}
The extremum of entropy as a function of filling is reported also in Refs.~\onlinecite{jarrellNFL,khatami,shila}. 
The physical origin of a peak of entropy as a function of doping will be discussed in Sec.~\ref{Sec:Plaquette}.
Here we can anticipate that this feature results from reorganization of the basic electronic excitations of the systems across the OD-UD transition.

\section{Phases characterization: cluster quantities}
\label{Sec:Phases}

Our results show that the normal-state phase diagram of the two-dimensional Hubbard model is controlled by a first-order transition between a metal and either a Mott insulator or another metal that evolves continuously from the Mott insulating state.
The phases separated by this transition have the same symmetry but different thermodynamic properties, like density, compressibility, entropy, double occupancy.

In this Section we further characterize the nature of the sequence of phases that arises upon doping.
We discuss the cluster Green's function and the cluster self-energy that contain direct information about the effects of electronic correlations.
Our main finding is that critical behavior, as seen in a large scattering rate, originates from the spinodal surface $\mu_{c1}(U,T)$ that delimits the underdoped phase.
As the carrier concentration moves away from that transition, towards both low or high doping, coherent electronic behavior is recovered.

\subsection{Green's function}
\label{SubSec:GF}

\begin{figure}[!ht]
\centering{\includegraphics[width=0.95\linewidth,clip=]{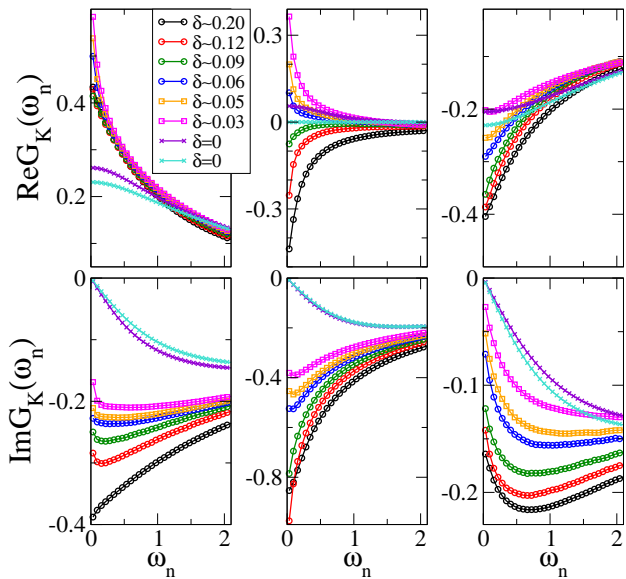}}
\caption{Real and imaginary parts of the cluster Green's function $G_{K}(i\omega_n)$ at $U=7.0t$ and low temperature $T/t=1/100$ for several dopings. Left panels show the results for the orbital $K=(0,0)$, central panels for $K=(\pi,0)$ and right panels for $K=(\pi,\pi)$. The finite (zero) value of the imaginary part of the cluster Green's function at $\omega_n\rightarrow 0$ indicates the metallic (insulating) character of the solution. Accordingly, three regimes can be distinguished: metal (circles), insulator (crosses) and a strong cluster momentum differentiation, with orbital $K=(\pi,\pi)$ gapped and $K=(0,0)$, $(\pi,0)$ gapless. The first and the last behavior characterize respectively the overdoped and the underdoped phases. The dopings on the figure correspond to $\mu=-2.0,-1.5,-1.3,-1.2,-1.15,-1.05,-0.2,0.0$ 
}
\label{fig5}
\end{figure}

We begin with a specific example. Fig.~\ref{fig5} shows the real and imaginary parts of the cluster Green's function for several dopings at $U=7t$, above $U_{\rm MIT}$, and at low temperature $T/t=1/100$.
As discussed in Sec.~\ref{Sec:Methodology}, in the $2\times2$ plaquette scheme the cluster Green's function in cluster momentum basis is a diagonal matrix with diagonal entries $G_{(0,0)}$, $G_{(0,\pi)}$, $G_{(\pi,0)}$, $G_{(\pi,\pi)}$.
These elements can be thought to represent a coarse grained average of the lattice Green's function over part of the Brillouin zone. The curves display a systematic evolution as the Mott insulating state is approached.

Let us analyze the low frequency behavior of the real and imaginary parts of the cluster Green's function.
The extrapolated value of Im$G_{K}(\omega\rightarrow 0)$  is an estimate of the local density of states at the Fermi level, a quantity accessible in photoemission experiments.
For all dopings shown, the $(\pi,0)$ component carries most of the spectral weight at the Fermi level.
At large values of doping, the imaginary part is finite in all orbitals (orbitals refer to $K$ values), showing that the system is metallic (see curves with circles).
This behavior distinguishes the overdoped phase.
On the other hand, at zero doping, the imaginary part goes to zero for all orbitals, so it indicates insulating Mott behavior for all cluster momenta (see curves with crosses).
In between these two regions, the imaginary part remains finite in the orbital $(\pi,0)$ and $(0,0)$, while it shows a clear tendency to go to zero in orbital $(\pi,\pi)$ (see curves with square symbols).
This implies that the former two orbitals are metallic and the latter is insulating, in striking contrast with the OD and non-interacting cases.
This phenomenon of strong momentum space differentiation~\cite{michelEPL,michelPRB,gullPRB,michelCFR} characterizes the UD phase and has a transparent physical interpretation: doping the Mott insulator occurs gradually in certain cluster momenta, leading to emergence of metallic behavior in some regions of the Brillouin zone but not in others.
Physically, this feature is the coarse-grained manifestation of the continuous appearance of the Fermi surface out of the Mott insulating state.
In the $2\times2$ plaquette scheme, the cluster momentum differentiation in the UD phase found in the low frequency behavior of Im$G_K(\omega_n)$ is compatible with arc or pocket formation on the Fermi surface.
Distinguishing between these scenarios falls outside the scope of the present study.
One has  to rely on periodization schemes to obtain the lattice Green's function~\cite{stanescuK,st,sakaiPRL,sakaiPRB}.
Another possibility, recently explored in the context of DCA calculations, is to consider progressively larger clusters~\cite{michelCFR,werner8,gullPRB}.
Other band structure effects not included here can also be important~\cite{st}.

Let's now study the real part of the Green's function, displayed in the top panels of Fig.~\ref{fig5}.
Amongst other things, this quantity encodes information about the particle-hole symmetry of the orbital at low frequency as follows from the spectral representation of the Green's function,
\begin{equation}
G(i\omega_n)= -\int \frac{d\omega}{\pi} \frac{ImG(\omega)}{i\omega_n-\omega}.
\end{equation}
In all the doping range, Re$G_{(0,0)}(\omega\rightarrow 0)$ is positive, implying that the largest spectral weight at low frequency lies below the Fermi level.
The opposite sign is found for the orbital $(\pi,\pi)$, meaning that it has most of the low frequency spectral weight above the Fermi energy.
In sharp contrast, the plot of Re$G_{(\pi,0)}(i\omega_n)$ reveals a change of sign at a characteristic doping.
This means particle-hole symmetry of the $(\pi,0)$ orbital at that doping, which is remarkably close to the doping where the above mentioned strong momentum space differentiation in ImG$_{(\pi,0)}(i\omega_n)$ sets in.
We shall see in Sec.~\ref{Sec:Discussion} that this property is closely associated with the large scattering rate found near the critical transition line.

\begin{figure*}[!ht]
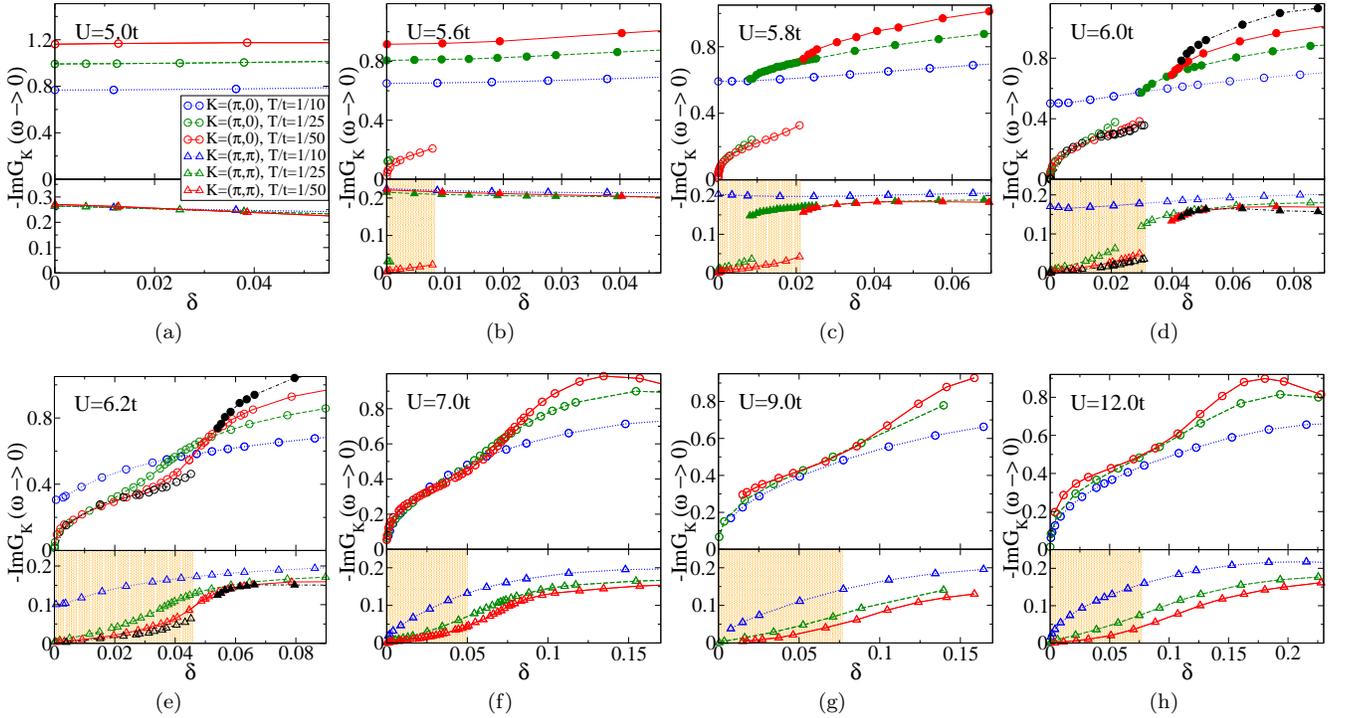

\centering
\subfigure[]{\label{fig6a}\includegraphics[width=0.24\linewidth]{fig6a.eps}}
\subfigure[]{\label{fig6b}\includegraphics[width=0.24\linewidth]{fig6b.eps}}
\subfigure[]{\label{fig6c}\includegraphics[width=0.24\linewidth]{fig6c.eps}}
\subfigure[]{\label{fig6d}\includegraphics[width=0.24\linewidth]{fig6d.eps}}

\subfigure[]{\label{fig6e}\includegraphics[width=0.24\linewidth]{fig6e.eps}}
\subfigure[]{\label{fig6f}\includegraphics[width=0.24\linewidth]{fig6f.eps}}
\subfigure[]{\label{fig6g}\includegraphics[width=0.24\linewidth]{fig6g.eps}}
\subfigure[]{\label{fig6h}\includegraphics[width=0.24\linewidth]{fig6h.eps}}
\caption{Extrapolated zero frequency value of the imaginary part of the cluster Green's function, -Im$G_{K}(\omega\rightarrow 0)$, as a function of doping $\delta=1-n$, for several values of the interaction strength $U$. For each figure, the upper panel shows cluster momentum  $K=(\pi,0)$ (circles) and the lower panel displays $K=(\pi,\pi)$ (triangles). Note the difference of scale of the y-axis. The data shown are for temperatures $T/t=1/10$ (blue dotted), $1/25$ (green dashed), $1/50$ (red solid), $1/100$ (black dot-dashed). When two solutions are found to coexist, the solutions obtained following the metallic and the insulating solution are indicated as full and open symbols respectively.
This observable measures the density of states at the Fermi energy averaged in a coarse-grained cluster momentum region.
Doping of the Mott state occurs gradually in cluster momentum, with the metallization that starts first in the orbital $K=(\pi,0)$ and $(0,0)$ while $K=(\pi,\pi)$ remains insulating. This behavior characterizes the underdoped phase and is highlighted by the gray (orange) background for the lowest temperature displayed. The transition between the underdoped and overdoped phase can be first order or it can be a crossover depending on $U$ and $T$.  In the latter case, for concreteness we define the boundary, illustrated in the figure for $T/t=1/50$, by -Im$G_{(\pi,\pi)}(\omega\rightarrow 0)=0.05$.
}
\label{fig6}
\end{figure*}
\begin{figure*}[!ht]
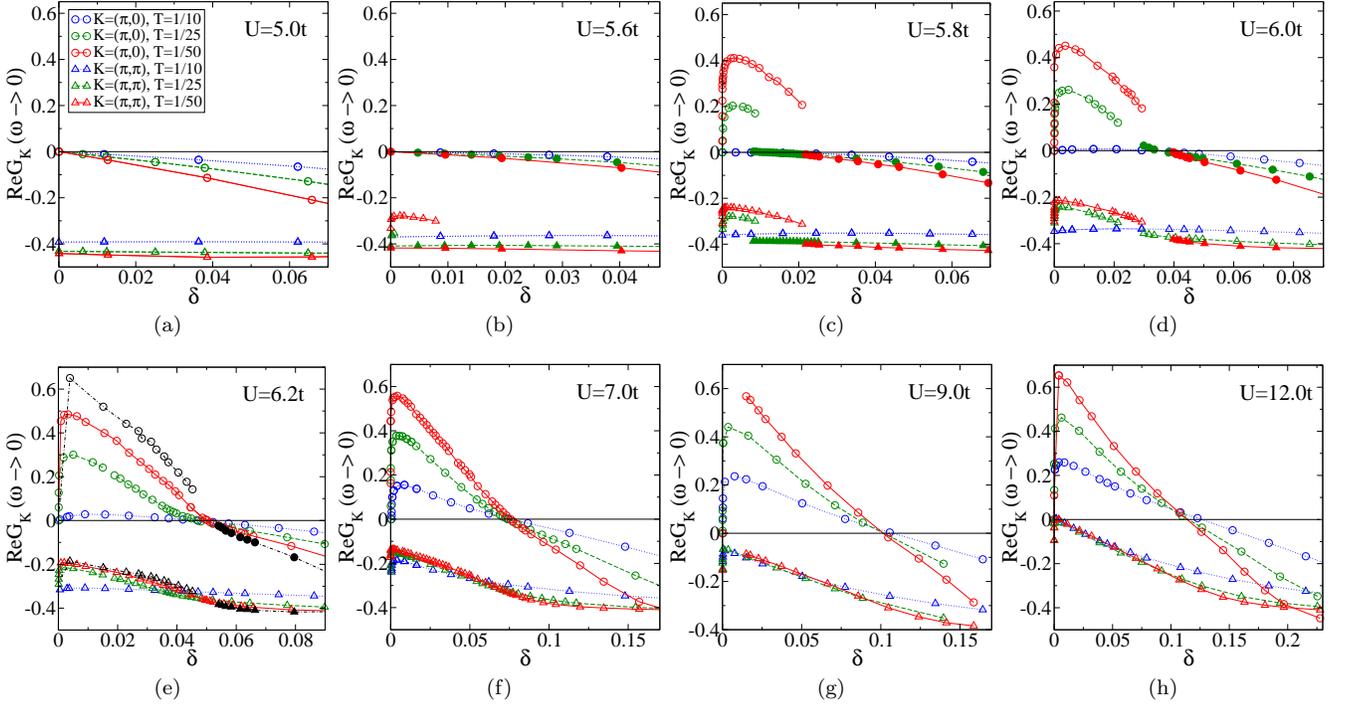

\centering
\subfigure[]{\label{fig7a}\includegraphics[width=0.24\linewidth]{fig7a.eps}}
\subfigure[]{\label{fig7b}\includegraphics[width=0.24\linewidth]{fig7b.eps}}
\subfigure[]{\label{fig7c}\includegraphics[width=0.24\linewidth]{fig7c.eps}}
\subfigure[]{\label{fig7d}\includegraphics[width=0.24\linewidth]{fig7d.eps}}

\subfigure[]{\label{fig7e}\includegraphics[width=0.24\linewidth]{fig7e.eps}}
\subfigure[]{\label{fig7f}\includegraphics[width=0.24\linewidth]{fig7f.eps}}
\subfigure[]{\label{fig7g}\includegraphics[width=0.24\linewidth]{fig7g.eps}}
\subfigure[]{\label{fig7h}\includegraphics[width=0.24\linewidth]{fig7h.eps}}
\caption{Re$G_K(\omega\rightarrow 0)$ as a function of doping $\delta$, for several values of the interaction strength $U$. Circles indicate the data for the $K=(\pi,0)$ orbital, triangles for the $K=(\pi,\pi)$. The data shown are for temperatures $T/t=1/10$ (blue dotted), $1/25$ (green dashed), $1/50$ (red circles), $1/100$ (black dot-dashed). When two solutions are found to coexist, the solutions obtained following the metallic and the insulating solution are indicated as full and open symbols respectively.
The assignment of symbols is the same as in Fig.~\ref{fig6}. This observable measures the low frequency asymmetry of the orbitals. The $(\pi,0)$ orbital crosses zero at a characteristic doping close to the underdoped-overdoped transition.
}
\label{fig7}
\end{figure*}

So far we have given a qualitative overview of our data for a single value of $U$ and a single low temperature that is nevertheless higher than first-order transition line $T_{\rm cr}$.
This set of results points to the following picture.
Metallic behavior in all orbitals characterizes overdoped phase.
Strong modulation of spectral weight at the Fermi level, where $(\pi,0)$ and $(0,0)$ orbitals are metallic and $(\pi,\pi)$ is gapped or damped, signals the onset of the underdoped phase.
In between these two phases, the $(\pi,0)$ orbital becomes strikingly symmetric at low frequency.

To understand if these are genuine properties of the phase diagram, we compile the zero-frequency extrapolation of the imaginary and real parts of the cluster Green's function for a large range of dopings and temperatures.
We first discuss the behavior of Im$G_K(\omega\rightarrow 0)$ as a function of doping, shown in Figs.~\ref{fig6}(a)-(h) for different values of $U$.
To emphasize the cluster momentum differentiation, each figure has two panels containing data for the orbital $K=(\pi,0)$ (curves with circles in the upper panel) and $(\pi,\pi)$ (curves with triangles in the lower panel).
The $(0,0)$ orbital behaves similarly to the $(\pi,0)$ one and is not shown.
The $(\pi,0)$ orbital contains most of the spectral weight at the Fermi level and shows the largest temperature variation.
Below the second-order critical line, two solutions of the cellular DMFT equations coexist for each orbital $K$.
In the overdoped phase Im$G_K(\omega\rightarrow 0)$ goes to a finite value for all orbitals $K$, revealing the metallic nature of this phase.
On the other hand, in the underdoped phase the $(\pi,0)$ and $(0,0)$ components of Im$G_K(\omega\rightarrow 0)$ go to a sizable value, except asymptotically close to zero doping, while the $(\pi,\pi)$ orbital remains basically gapped or with a very small value.
The gray (orange) background in the lower panels highlights this region for the lowest temperature shown.
At the spinodal surfaces $\mu_{c1}(T,U)$ and $\mu_{c2}(T,U)$  these two solutions exhibit an abrupt jump in each orbitals: from essentially zero to a finite value in the case of $K=(\pi,\pi)$, and from a finite value to a larger value in the case of $K=(\pi,0)$ and $(0,0)$ (see $U/t=5.6, 5.8$ and $6.0$).
Above the critical end line the two solutions merge and the transition between the two phases occurs as a crossover.

Since $T_{\rm cr}$ decreases with increasing $U$, an immediate implication of our results is the following: a study focusing {\it only} on a large interaction strength $U$ will detect the changes in Im$G_K(\omega\rightarrow 0)$, but not the underlying transition between two metals at finite doping, due to the low energy scales involved (see $U/t=7.0,9.0,12.0$).
On the contrary, our exhaustive scan of the 3D phase diagram allows us to show a direct link between the momentum differentiation and the finite doping first-order transition.

We note that in the UD phase, Im$G_{(\pi,\pi)}(\omega\rightarrow 0)$ does not actually seem to extrapolate exactly to zero, but rather to a vanishingly small value.
Even if the extrapolated value decreases as $T$ is lowered, and turns from sublinear to overlinear dependence on doping, we rationalize this feature by the fact that in cellular DMFT the orbitals $K$ are coupled through both the Coulomb interaction $U$ and the hopping, making difficult a sharp decoupling among the orbitals at finite temperature.
The latter behavior is found for example in DCA calculations~\cite{michelEPL,michelPRB,werner8,gullPRB,michelCFR}, where the coupling among the orbitals occurs through the interaction $U$ only.

Fig.~\ref{fig7} shows the extrapolated values of the real part of the cluster Green's function, Re$G_K(\omega\rightarrow 0)$, a measure of the particle-hole symmetry of the orbitals at low frequency.
Here the most striking feature is the change of sign in the $(\pi,0)$ component that occurs, apart from the obvious particle-hole symmetric line $\mu=0$, at $\delta=0$, at the spinodal surface $\mu_{c2}(U,T)$ that signals the disappearance of the OD phase.
This feature is not restricted to temperatures below the critical line $T_{\rm cr}$ but also continues for temperatures above it.

\subsection{Self-energy}
\label{SubSec:Sigma}

Precious information about the nature of the two phases separated by the first-order transition follows from the analysis of the cluster self-energy.
\begin{figure}[!ht]
\centering{\includegraphics[width=0.9\linewidth,clip=]{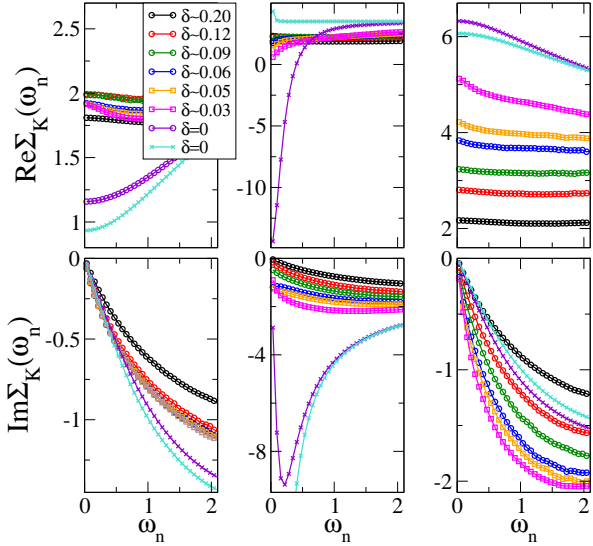}}
\caption{Real and imaginary part of the cluster self-energy $\Sigma_{K}(i\omega_n)$ at $U=7.0t$ and the low temperature $T/t=1/100$ for several dopings. Left panels show the results for the orbital $K=(0,0)$, central panels for $K=(\pi,0)$ and right panels for $K=(\pi,\pi)$. The assignment of symbols is the same as in Fig.~\ref{fig5}.
}
\label{fig8}
\end{figure}
In Fig.~\ref{fig8} we present a selection of self-energies for the same parameter values as in Fig.~\ref{fig5}.

An instructive way to look at the data is to consider the concept of Fermi liquid coherence {\it within the cluster momentum}.
In Landau's Fermi liquid theory the low energy excitations of an interacting system can be described as long-lived particles, called quasiparticles.
This Fermi liquid regime applies below a characteristic temperature, the coherence temperature, and a necessary condition for its existence is that the imaginary part of the self-energy goes to zero as $\omega\rightarrow 0$ with the form  $\Sigma\approx i\omega_nA +B$ with $A$ negative.

The self-energy of both the $(0,0)$ and $(\pi,\pi)$ orbitals in Fig.~\ref{fig8} obeys this relation for all the metallic states (circles and squares) of the doped Mott insulator, revealing Fermi liquid coherence of the electronic excitations.
Note that the linear low frequency behavior of Im$\Sigma_{(0,0)}$ and Im$\Sigma_{(\pi,\pi)}$ persists in the Mott insulating state (lines with crosses), where the particles are localized.
There the system is gapped (see the corresponding low frequency vanishing behavior of Im$G_{(0,0)}(\omega_n)$ and Im$G_{(\pi,\pi)}(\omega_n)$ in Fig.~\ref{fig5}).

In contrast, the imaginary part of the $(\pi,0)$ self-energy shows a non-monotonic behavior as a function of doping.
The metallic state at both large and small doping displays Fermi liquid coherence (Im$\Sigma_{(\pi,0)}(\omega_n\rightarrow 0)$ goes linearly to zero), but at intermediate doping, at the crossover between the underdoped-overdoped phases, the large finite intercept indicates a sudden drop of electronic coherence (see curves for $\delta=0.05,0.06$).
Finally, the Mott insulating state is reached and a clear pole appears in the imaginary part of the $(\pi,0)$ self-energy at $\delta=0$ and $\mu=0$.

\begin{figure*}[!ht]
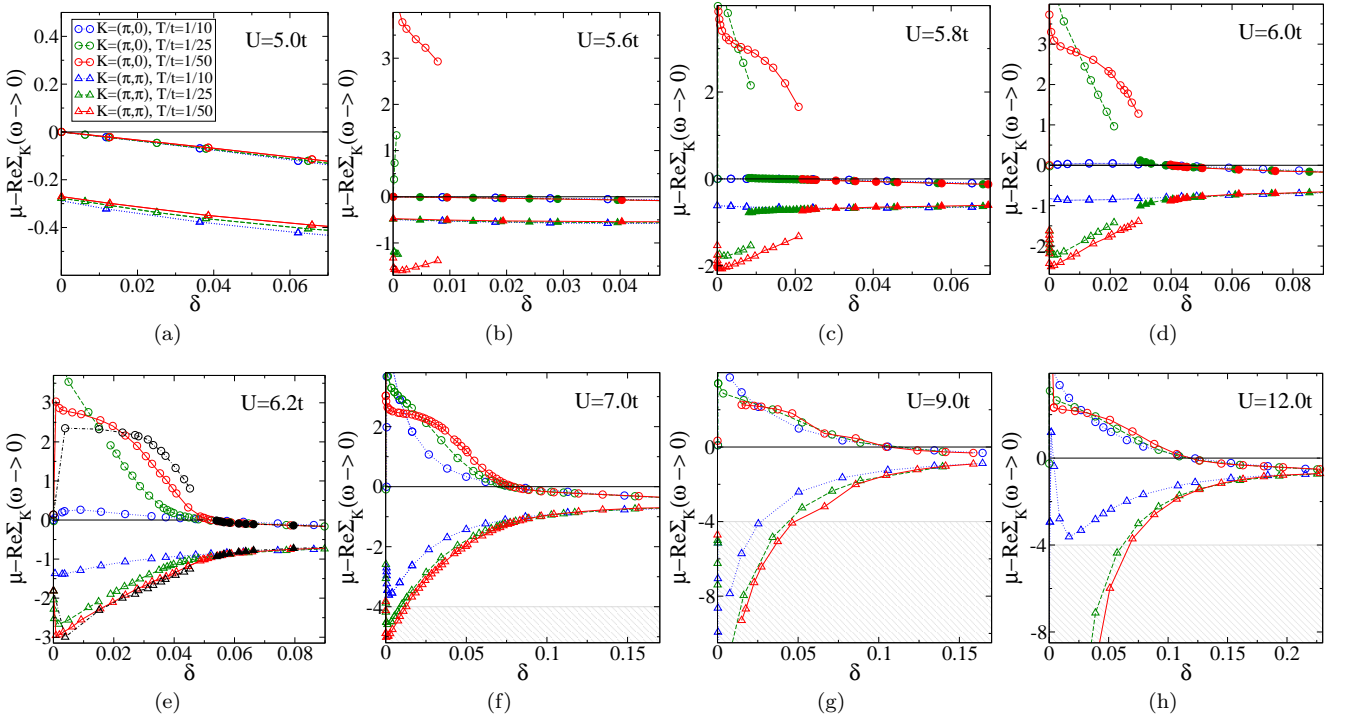

\centering
\subfigure[]{\label{fig9a}\includegraphics[width=0.24\linewidth]{fig9a.eps}}
\subfigure[]{\label{fig9b}\includegraphics[width=0.24\linewidth]{fig9b.eps}}
\subfigure[]{\label{fig9c}\includegraphics[width=0.24\linewidth]{fig9c.eps}}
\subfigure[]{\label{fig9d}\includegraphics[width=0.24\linewidth]{fig9d.eps}}

\subfigure[]{\label{fig9e}\includegraphics[width=0.24\linewidth]{fig9e.eps}}
\subfigure[]{\label{fig9f}\includegraphics[width=0.24\linewidth]{fig9f.eps}}
\subfigure[]{\label{fig9g}\includegraphics[width=0.24\linewidth]{fig9g.eps}}
\subfigure[]{\label{fig9h}\includegraphics[width=0.24\linewidth]{fig9h.eps}}
\caption{Effective chemical potential $\mu_{\rm eff}^{K}=\mu-Re\Sigma_K(\omega\rightarrow 0)$ renormalized by the electronic correlation versus $\delta$. Data are shown for the cluster momenta  $K=(\pi,0)$ (circles) and $K=(\pi,\pi)$ (triangles) and are obtained for temperatures $T/t=1/10$ (blue dotted line), $1/25$ (green dashed line), $1/50$ (red solid line), $1/100$ (black dot-dashed line). Note that $\mu_{\rm eff}^{(\pi,0)}$ crosses zero at a characteristic doping close to the OD-UD transition, or more precisely at the spinodal $\mu_{\rm c2}(U,T)$ and its high temperature crossover. 
The hatched grey lines indicate the region where $\mu_{\rm eff}^{(\pi,\pi)}$ exceeds the noninteracting bandwidth of the $(\pi,\pi)$ orbital. 
}
\label{fig9}
\end{figure*}
\begin{figure*}[!ht]
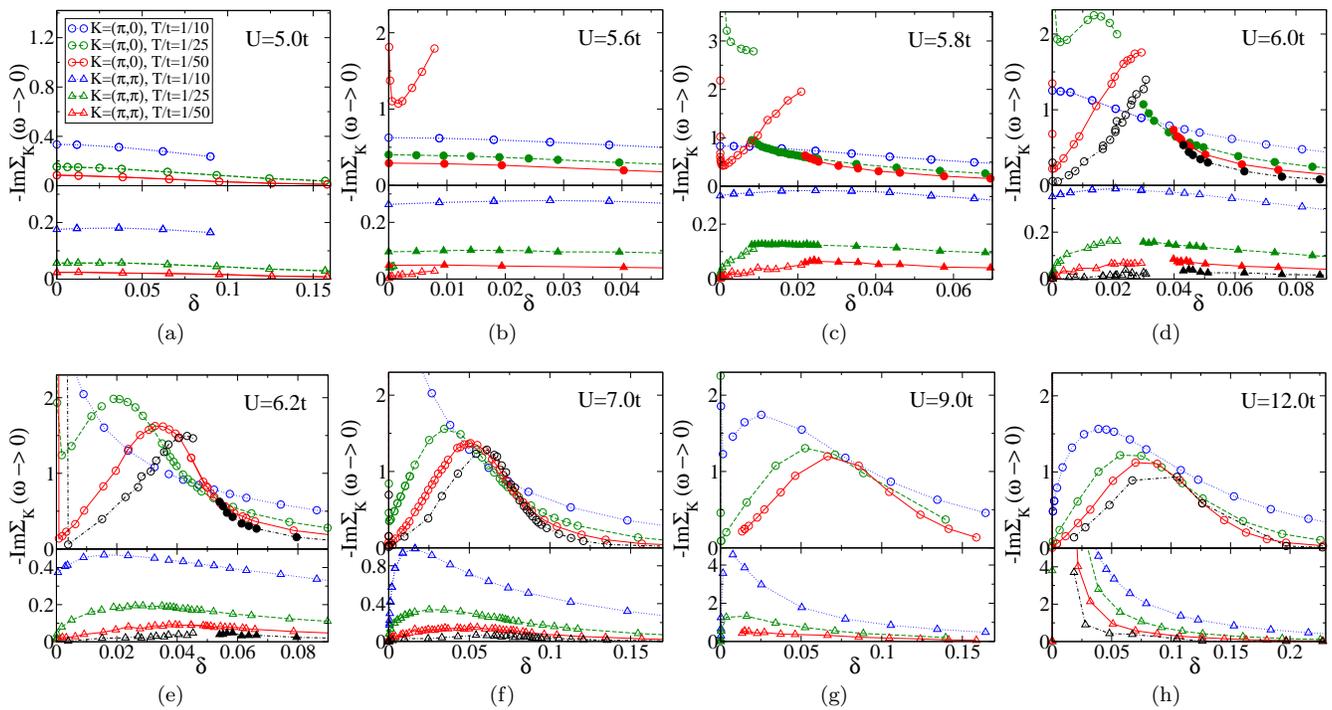

\centering
\subfigure[]{\label{fig10a}\includegraphics[width=0.24\linewidth]{fig10a.eps}}
\subfigure[]{\label{fig10b}\includegraphics[width=0.24\linewidth]{fig10b.eps}}
\subfigure[]{\label{fig10c}\includegraphics[width=0.24\linewidth]{fig10c.eps}}
\subfigure[]{\label{fig10d}\includegraphics[width=0.24\linewidth]{fig10d.eps}}

\subfigure[]{\label{fig10e}\includegraphics[width=0.24\linewidth]{fig10e.eps}}
\subfigure[]{\label{fig10f}\includegraphics[width=0.24\linewidth]{fig10f.eps}}
\subfigure[]{\label{fig10g}\includegraphics[width=0.24\linewidth]{fig10g.eps}}
\subfigure[]{\label{fig10h}\includegraphics[width=0.24\linewidth]{fig10h.eps}}
\caption{Extrapolated zero-frequency value of the imaginary part of the cluster self-energy, -Im$\Sigma_{K}(\omega\rightarrow 0)$ as a function of doping $\delta$ and for several values of the interaction $U$. For each figure, the upper panel shows cluster momentum  $K=(\pi,0)$ (circles) and the lower panel displays $K=(\pi,\pi)$ (triangles). This quantity is proportional to the scattering rate $\Gamma_{K}$. The data shown are for temperatures $T/t=1/10$ (blue dotted line), $1/25$ (green dashed line), $1/50$ (red solid line), $1/100$ (black dot-dashed line). -Im$\Sigma_{(\pi,0)}(\omega\rightarrow 0)$ is peaked close to the spinodal surface $\mu_{c1}(U,T)$ of the first-order transition and to its high temperature crossover. That peak reaches its overall maximum in the 3D phase diagram at the Mott end point $U_{\rm MIT}\approx 5.95$ and $\delta=0$ (see Figs.~\ref{fig10c},\ref{fig10d}). As the temperature increases, the value of -Im$\Sigma_{(\pi,0)}(\omega\rightarrow 0)$ at its maximum increases as it does its width in doping. The peak in -Im$\Sigma_{(\pi,0)}(\omega\rightarrow 0)$ then disappears at a characteristic temperature which is progressively higher as $U$ increases.
}
\label{fig10}
\end{figure*}

The main results of the above analysis are a buildup of incoherent metallic behavior close to the first-order transition and its high temperature crossover.
It is of key importance to investigate if this picture survives in the whole phase diagram.
To this end, we show in Fig.~\ref{fig9} the effective chemical potential, which corresponds to the chemical potential renormalized by correlation effects, and is defined by $\mu_{\rm eff}^{K}=\mu-Re\Sigma_K(\omega\rightarrow 0)$.  
In Fig.~\ref{fig10} we plot the scattering rate $\Gamma_{K}$, estimated from the  zero-frequency extrapolation of the imaginary part of the cluster self-energy $=-Im\Sigma_{K}(\omega\rightarrow 0)$. 
Although these quantities are not strictly equal, we will use them interchangeably.
We use a linear interpolation and we have verified that a quadratic interpolation gives qualitatively similar results.
For clarity we show results only for the orbitals $K=(\pi,0)$ (circles) and $(\pi,\pi)$ (triangles).

The most important result, shown in the upper panels of Fig.~\ref{fig10}, is a large scattering rate $-Im\Sigma_{(\pi,0)}(\omega\rightarrow 0)$ for a finite range of dopings peaked at the first-order transition between two types of metals.
A large scattering rate is in conflict with Fermi liquid theory and indicates that short-lived (i.e. incoherent) excitations, and not long-lived quasiparticle, produce the metallic behavior in that regime.
Nevertheless, if one goes away from the transition, the scattering rate rapidly drops to small values and thus crosses over to a conventional Fermi liquid with small intercept.
The overall behavior of $-Im\Sigma_{(\pi,0)}(\omega\rightarrow 0)$ unequivocally shows that the source of scattering comes from the transition between the underdoped and the overdoped phase.
Some recent studies, using different cluster methods, already reported a peak in the scattering rate at finite doping, associating it to a pseudogap phenomenon\cite{liebsch}, to a competing superexchange and Kondo scale\cite{hauleDOPING,michelEPL}, or to a selective metal-insulator transition\cite{michelEPL,michelCFR}. Our contribution is twofold: first, we link this feature to the spinodal $\delta_{c1}(T)$ and its high temperature precursor; second, by tracking this feature in the $(U-T-\delta)$ phase diagram, we show that it is connected to the Mott endpoint $U_{\rm MIT}$, hence unveiling its unexpected origin in the Mott physics. 

Let us study in detail the behavior of the self-energy in the two phases.
The {\it overdoped phase} has a rich behavior as doping is reduced towards the first-order transition surface.
At large doping, far from the transition, the effective chemical potential of all orbitals merges, as demonstrated in Fig.~\ref{fig9}.
This is clear at large $U$ (cf. Fig.~\ref{fig9}(g),(h)) and indeed occurs at smaller values of $U$ for similar dopings (not shown).
Hence, the self-energy is local (momentum independent) and the single-site DMFT solution captures the physics of the problem.
As the transition is approached, $\mu_{\rm eff}^K$ for different $K$ separate.
Similar trend has been found in DCA calculations for different sizes of cluster~\cite{michelCFR}.
The corresponding scattering rate $\Gamma_{K}$ in Fig.~\ref{fig10} is quite small, decreases with decreasing $T$ and monotonically increases as the doping is reduced towards the first-order transition surface.
Eventually, at the spinodal $\delta_{c2}(U,T)$, $\Sigma$ shows a sudden jump (Figs.~\ref{fig9},\ref{fig10}) and interestingly the effective chemical potential of the orbital that carries most of the spectral weight at the Fermi level, $\mu_{\rm eff}^{(\pi,0)}$, changes sign.
The overall low frequency behavior of Im$\Sigma_K$ within the overdoped phase indicates that the system is a strongly interacting Fermi liquid and the effects of electronic correlations dramatically increase as the first-order transition or its high temperature crossover are approached.

Let's now turn to the {\it underdoped phase} that originates from the Mott insulator.
The analysis of the cluster Green's function revealed that the $(\pi,\pi)$ orbital is gapped but the other orbitals are gapless.
Close to the Mott insulator and at low enough temperature, the scattering rate $\Gamma_{(\pi,0)}$ is small and decreases as $T$ decreases.
As the the boundary with the first-order transition to the overdoped phase is approached, $\Gamma_{(\pi,0)}$ increases and eventually reaches its maximum close to the spinodal surface $\mu_{c1}(U,T)$ or near the crossover line above the critical line.
A further confirmation of this picture emerges from the striking temperature evolution of $\Gamma_{(\pi,0)}$ displayed in Fig.~\ref{fig10}. As the temperature increases, the value of $\Gamma_{(\pi,0)}$ at its maximum increases as does its width in doping.
Interestingly, the lower is $T_{\rm cr}$ the higher is the temperature range at which the value of the $(\pi,0)$ scattering rate displays a maximum with doping.
Note that the overall maximum as a function of $U$ peaks around $U_{\rm MIT}$, relating directly this phenomenon to Mott physics.

In contrast, the $(\pi,\pi)$ orbital is insulating and therefore $\Gamma_{(\pi,\pi)}$ looses its usual meaning of a scattering rate.
In search for a possible explanation, we note that, in the temperature range explored, $\mu_{\rm eff}^{(\pi,\pi)}$ has a large value, that can even exceed the noninteracting bandwidth for sufficient large $U$, as marked by the hatched grey lines in Figs.~\ref{fig9}~f,g,h.
It has also been discussed, in the context of DCA framework and for $U=7t$, that a pole in the self-energy is responsible for the insulating behavior~\cite{werner8,gullPRB,michelCFR}.
Our cellular DMFT results for $U=12.0t$ can be compatible with the existence of that pole, however for smaller values of $U$ that pole seems not to be present or is strongly reduced, as was already pointed out in Ref.~\onlinecite{hauleDOPING}.
Even if this mechanism remains to be better clarified, our contribution is to track the origin of the large scattering rate to the spinodal surface $\mu_{c1}(T,U)$ and its high temperature crossover.

Finally, we stress that the coherence of the $(\pi,0)$ orbital in the underdoped phase at low doping, as signaled by the vanishing of the corresponding scattering rate in Fig.~\ref{fig10}, does not imply that the system as a whole is a Fermi liquid.
As discussed before, the $(\pi,\pi)$ orbital remains gapped in this phase suggesting a ``small'' Fermi surface compatible with arc or pocket formation. 
In other words, it is more likely that in the underdoped phase electronic coherence does not result from quasiparticle propagation but from another mechanism yet to be identified. In the next section we shall argue that coherence behavior is associated to hole doped carriers moving in a background of spin singlets formed by the superexchange mechanism.

What picture emerges from the analysis of cluster quantities?
Breakdown of Fermi liquid behavior, revealed by a dramatic enhancement of scattering rate  $\Gamma_{(\pi,0)}$, is realized over a large region of the phase diagram close to the first-order transition between the underdoped and overdoped metals and reaches its overall maximum at the Mott endpoint $U_{\rm MIT}$.
Thus our results pin down that incoherent states come from Mott physics, even for regions way beyond half filling.
As we move away from the transition, the lifetime of the low energy excitations of the metallic state sharply increases.

Fermi liquid breakdown beyond half filling is unambiguously tied to Mott physics.
The last piece of this puzzling scenario is to identify the physical mechanism driving this unconventional behavior.
This is the focus of the next section.

\section{Role of short-range spin correlations}
\label{Sec:Plaquette}

To shed light on the physical origin of the incoherent metallic behavior associated with the first-order transition, we now turn to the analysis of the short-range spin correlations.
In fact, cellular DMFT takes into account on equal footing local quantum fluctuations and finite length spatial fluctuations (within the cluster size, here a $2\times2$ plaquette).
The inclusion of spatial correlations manifests itself mathematically in the non locality of the self-energy~\cite{gabiCDMFT,kotliarRMP}.
In this section we first compare our cellular DMFT phase diagram with that of single-site DMFT, where magnetic correlations in space are absent.
The differences between the two phase diagrams directly reveals the effects of the short-range correlations.
Then we characterize the phases that occur across the transition in terms of these fluctuations.
As we shall see, in the Mott insulating state the electrons are localized due to strong screened Coulomb interaction and their spins lock into singlets states due to the superexchange mechanism.
In the UD phase, i.e. the metallic state that evolves out of the Mott insulator, the system is still dominated by singlet states and metallic behavior is associated with doped charge carriers that propagate in a background of spins with strong singlet correlations due to superexchange.
Finally, in the OD phase, the basic excitations of the system are quasiparticles due to conventional Kondo screening. There are analogies with the suggestion of Ref.~\onlinecite{hauleDOPING}.

\subsection{Comparison with the single-site DMFT phase diagram}
\label{SubSec:DMFT}
A natural way to understand the effects of the short-range spin correlations on the Mott transition is to compare our cellular DMFT phase diagram, Fig.\ref{fig2}, that accounts for these magnetic correlations, with that of single-site DMFT, where those correlations are absent.

Sustained investigations have firmly set the single-site scenario for the Mott transition in the single band Hubbard model~\cite{mzk,mkz,zrk,antoine,sahana,mck,wernerD,rmp,phytoday} and the resulting phase diagram as a function of chemical potential, interaction strength and temperature can be found for instance in Ref.~\onlinecite{sahana}.
Here the key phenomenon that governs the whole phase diagram is a finite temperature first-order transition originating at the finite temperature Mott critical point.
This transition occurs between a Mott insulator and a correlated metallic state, and can take place as a function of either $U$, $T$ or $\mu$.

Similarly to the single-site DMFT case, a first-order transition dominates the normal-state phase diagram also in the cellular DMFT case, as we demonstrated in the previous sections.
But the nature of this first-order transition is different and the fundamental difference comes from short-range spin correlations considered in cellular DMFT.

Three main effects can be identified.
First, they reduce the value of the Mott endpoint $U_{\rm MIT}$.
This is as expected because the threshold to open a Mott gap increases as the frustration at short-distance increases \cite{Kyung:2006,andriy,phk}.

Second, the short-range magnetic correlations change the shape of the spinodal lines~\cite{phk,sht}.
In single-site DMFT the boundary of the first-order transition bends towards the correlated metallic state (as a function of either the interaction strength $U$ or the chemical potential $\mu$), so that the insulating phase can be reached upon increasing the temperature of the system.
According to the Clausius-Clapeyron equation this comes from the fact that the correlated metallic state has lower entropy than the Mott insulating state.
This behavior occurs in several correlated systems, like close to the first-order transition line in $V_2O_3$~\cite{mcwhan,limelette}. 
Physically, this is because in taking the large lattice connectivity limit to justify single-site DMFT, the magnetic exchange coupling between two neighboring spins drops out, thus resulting in a paramagnetic Mott insulator with large spin degeneracy of order $N\log2$ (where $N$ is the number of lattice site)~\cite{rmp}.
As discussed in section~\ref{Sec:Thermodynamics}, the opposite behavior occurs in cellular DMFT, where the first-order transition bends towards the Mott insulating state.
This reveals that the UD state (and its parent Mott insulating phase) has lower entropy than the OD metallic state as a consequence of short-range singlet formation that lifts the $2^N$ degeneracy due to the spins of the localized electrons~\cite{phk,balzer}.
The shape of the phase boundary of cellular DMFT at half-filling is observed in low-dimensional correlated systems, such as the organic conductors~\cite{Lefebvre:2000,kagawaPRB}.

The third and arguably the most surprising effect brought about by the short-range magnetic correlations is the mere appearance of the UD phase in the cellular DMFT phase diagram.
In both single-site and cellular DMFT solutions, the $\mu_{c2}(U,T)$ surface marks the vanishing of a correlated metallic solution.
On the other hand, the spinodal $\mu_{c1}(U,T)$ always marks the vanishing of the insulating solution within single-site DMFT, whereas --as we demonstrated in Sec.~\ref{Sec:Transition}--, this is generally not the case within cellular DMFT.
Doping instead proceeds gradually in certain cluster momenta. We stress that the apparent gradual doping of the Mott insulator observed over an extremely narrow range of dopings (~$0.002$, i.e. at least $10$ times smaller than what we find) in single-site DMFT is purely a finite temperature effect. 
In our case, short-range magnetic correlations do have the effect of shifting the spinodal surface $\mu_{c1}(U,T)$ to {\it finite} values of doping, implying that the first order transition can take place between two metallic phases (the UD and OD phases) instead of between an insulator and a metal.
Thus the UD phase occurs only in the cluster DMFT case and traces its roots to the short range-singlet formation.

\subsection{Plaquette eigenstates}
\label{SubSec:plaquette}

Having established the basic effects of the short range-spatial correlations on the structure of the phase diagram, we now concentrate in the relative importance of various many body states in the cellular DMFT solution of the Hubbard model.
To address this issue we scrutinize the relative statistical weight of the eigenstates $|m\rangle$ of the plaquette Hamiltonian $H_{\rm plaquette}$.
In the context of DMFT calculations, the analysis of the eigenstates of the quantum impurity problem has proven to lead to useful insights on the role of short-range spatial magnetic correlations~\cite{hauleCTQMC,michelPRB,hauleDOPING,phk,gullEPL,michelEPL} and also helped to construct a physical picture of the nature of the doped carriers~\cite{hauleDOPING,shim:nature}.

Cellular DMFT maps the lattice onto a $2\times2$ cluster of sites immersed in a self-consistent bath.
The quantum impurity problem defined by Eq.~(\ref{eq:action}) has a simple physical interpretation: as a function of imaginary time, the plaquette undergoes transitions between the $4^4$ possible quantum states by exchanging electrons with the rest of the lattice represented by the surrounding bath.
The bath hybridization function $\hat{\Delta}(\omega)$ encodes the dynamics of these processes.
It is possible to compute the statistical weight $P_{m}$ of the eigenstates ${|m\rangle}$ of the plaquette (see Refs.~\onlinecite{hauleCTQMC} for a rigorous definition). 
It can be interpreted as the relative time that the plaquette spends in the cluster eigenstate $|m\rangle$ and is the quantity we focus on.

The plaquette eigenstates $|m\rangle$ can be labeled by their number of electrons $N$, their total spin $S$ and their cluster momentum $K$ and physically can be thought of as representing a coarse grained approximation of the many body excitations of the system.

\begin{figure}[!ht]
\centering{\includegraphics[width=0.7\linewidth,clip=1]{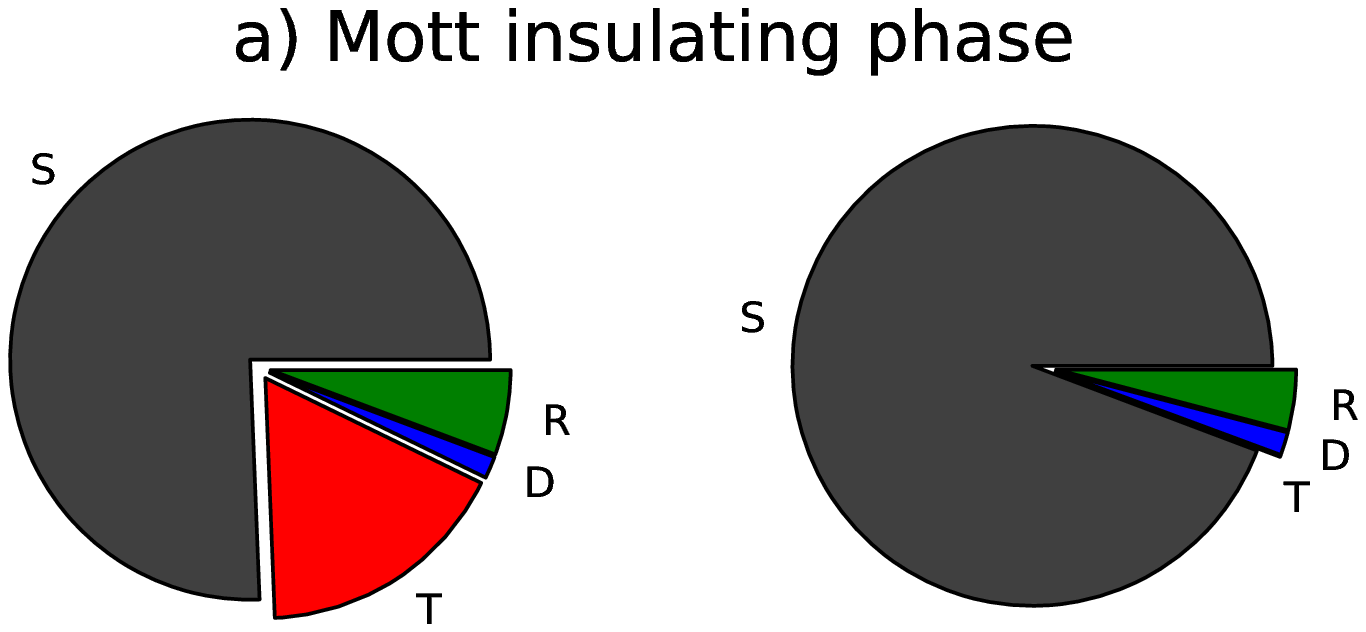}}
\centering{\includegraphics[width=0.7\linewidth,clip=1]{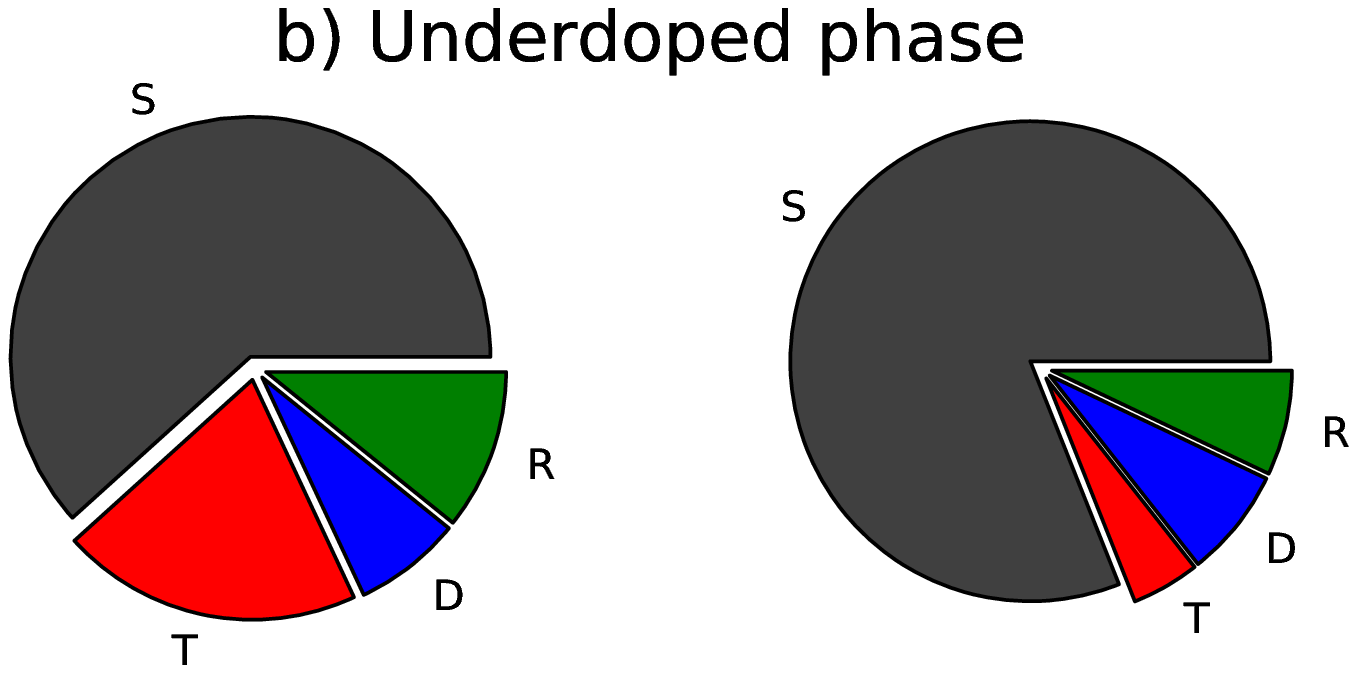}}
\centering{\includegraphics[width=0.7\linewidth,clip=1]{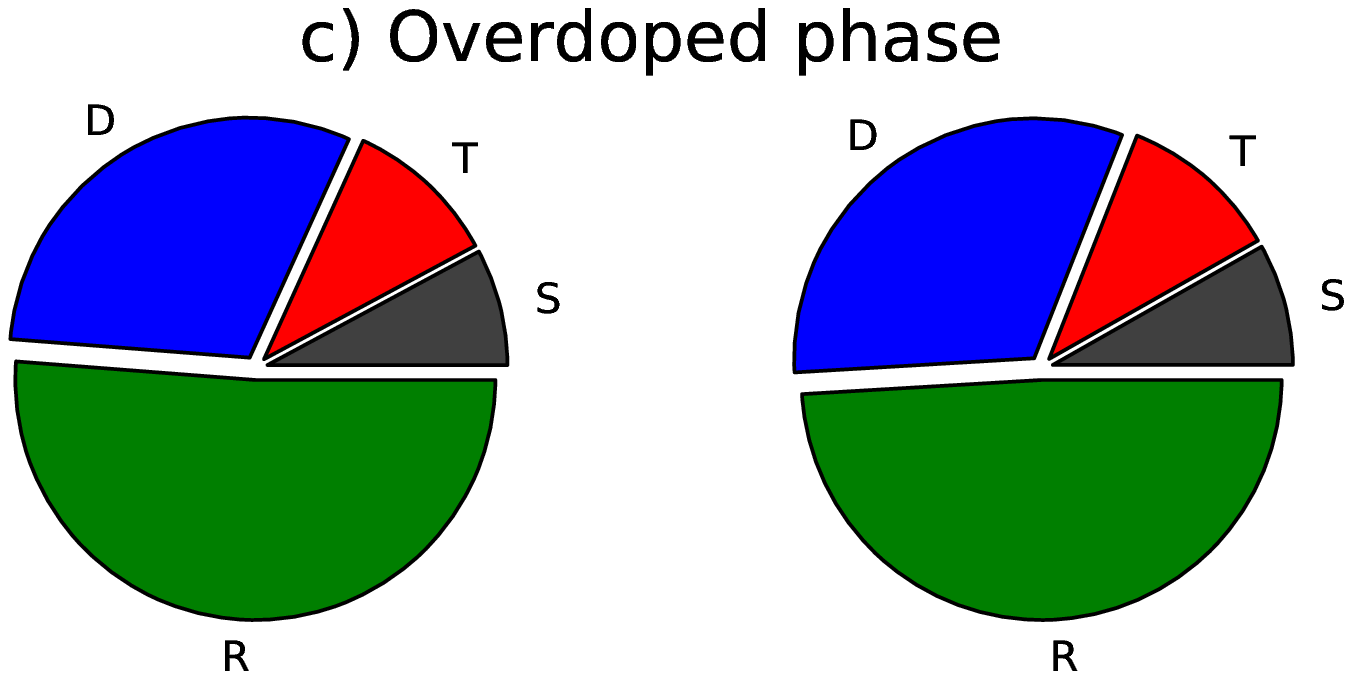}}
\caption{Statistical weight $P$ of the following cluster eigenstates: the singlet $S$ $|N=4; S=0; K=(0,0) \rangle$, triplet $T$ $|N=4; S=1; K=(\pi,\pi)\rangle$, and doublet $D$ $|N=3; S=1/2; K=(\pi,0)\rangle$, where $N$, $S$, and $K$ are the number of electrons, the total spin, and the cluster momentum of the cluster eigenstate. The remnant statistical weight is summed up in the histogram $R$. Data are obtained at $U=12.0t$ and shown at $T/1=1/10$ (left) and $1/50$ (right). The value of doping is (a) $\delta=0$, representative of the Mott insulating phase. (b) $\delta=0.02$, corresponding to the UD phase and (c) $\delta=0.25$, in the OD phase.}
\label{fig11}
\end{figure}
In Fig.~\ref{fig11} we plot the probability of the plaquette eigenstates for $U=12.0t$ for values representative of the Mott insulating state (Fig.~\ref{fig11}(a)), the UD phase (Fig.~\ref{fig11}(b)), and the OD phase (Fig.~\ref{fig11}(c)).
Results are shown for the high temperature $T/t=1/10$ and the low temperature $T/t=1/50$ (left and right respectively).
Focusing on the region of parameters close to the first-order transition, we find that there are only a few plaquette eigenstates with large probability.
They are the singlet with four electrons in the cluster momentum $K=(0,0)$ (black portions of the diagram):
\begin{equation}
| N=4, S=0, K=(0,0) \rangle,
\end{equation}
the triplet with four electrons and $K=(\pi,\pi)$ (red portion of the diagrams):
\begin{equation}
| N=4, S=1, K=(\pi,\pi) \rangle,
\end{equation}
and the doublet with three electrons and $K=(\pi,0)$ and the degenerate $K=(0,\pi)$ (blue portion of the diagrams).
\begin{equation}
| N=3, S=1/2, K=(\pi,0) \rangle.
\end{equation}
We sum up the probability of the remnant states (green portion of the diagrams).

The distribution of the statistical weight in the Mott insulator and UD phase is similar as does its evolution with temperature. This is expected since the UD phase evolves out of the mother Mott insulating phase. Let us now summarize the three main differences between the two metallic phases.
First, in the UD phase the plaquette electrons are locked into one prevailing configuration (half-filled singlet), whereas in the OD phase they fluctuate among several states.
Second, the temperature evolution of the plaquette eigenstates qualitatively differs in the two phases. While charge excitations are weakly $T$ dependent in both phases, the basic effect of decreasing the temperature strongly suppresses the spin triplet excitations in favor of the spin singlet ones, especially in the UD phase.
This has to be associated with the $T$ dependence of the local density of states at Fermi level shown in Fig.~\ref{fig6}.
Third, the mechanism behind coherence is also qualitatively different in the two phases.
In the OD phase we associate the onset of coherence to standard Kondo screening of the conduction electrons of the bath in which the plaquette is immersed.
In contrast, coherent propagation in the UD phase occurs when doped holes move in a magnetically well defined environment characterized by spins bounded into short-range singlet correlations via the antiferromagnetic superexchange mechanism.

\begin{figure*}[!ht]
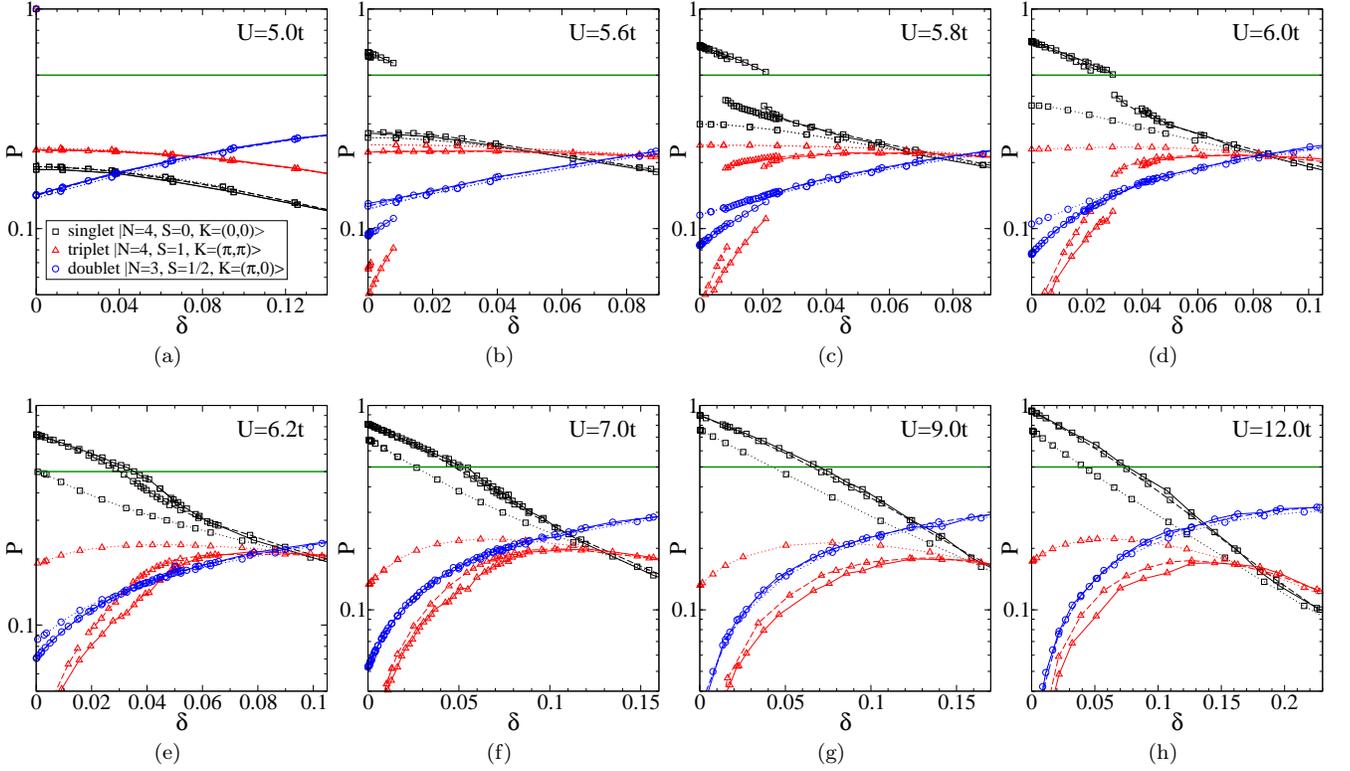

\centering
\subfigure[]{\label{fig12a}\includegraphics[width=0.24\linewidth]{fig12a.eps}}
\subfigure[]{\label{fig12b}\includegraphics[width=0.24\linewidth]{fig12b.eps}}
\subfigure[]{\label{fig12c}\includegraphics[width=0.24\linewidth]{fig12c.eps}}
\subfigure[]{\label{fig12d}\includegraphics[width=0.24\linewidth]{fig12d.eps}}

\subfigure[]{\label{fig12e}\includegraphics[width=0.24\linewidth]{fig12e.eps}}
\subfigure[]{\label{fig12f}\includegraphics[width=0.24\linewidth]{fig12f.eps}}
\subfigure[]{\label{fig12g}\includegraphics[width=0.24\linewidth]{fig12g.eps}}
\subfigure[]{\label{fig12h}\includegraphics[width=0.24\linewidth]{fig12h.eps}}
\caption{Statistical weight $P$ of the following cluster eigenstates as a function of doping $\delta$, for several values of $U$: the singlet $|N=4, S=0, K=(0,0)\rangle$, the triplet $|N=4, S=1, K=(\pi,\pi)\rangle$,  and the doublet $|N=3, S=1/2, K=(\pi,0)\rangle$ (black squares, red triangles, and blue circles respectively) where $N$, $S$ and $K$ are the number of electrons, the total spin and the cluster momentum of the cluster eigenstate. The data shown are for temperatures $T/t=1/10, 1/25, 1/50$ (dotted, dashed and solid line respectively).
The statistical weight $P$ of the half-filled singlet crosses $1/2$ (green solid horizontal line) roughly at the transition between the UD and OD phase, suggesting competing singlet spin versus other spins and charge fluctuations at the transition.
}
\label{fig12}
\end{figure*}
\begin{figure*}[!ht]
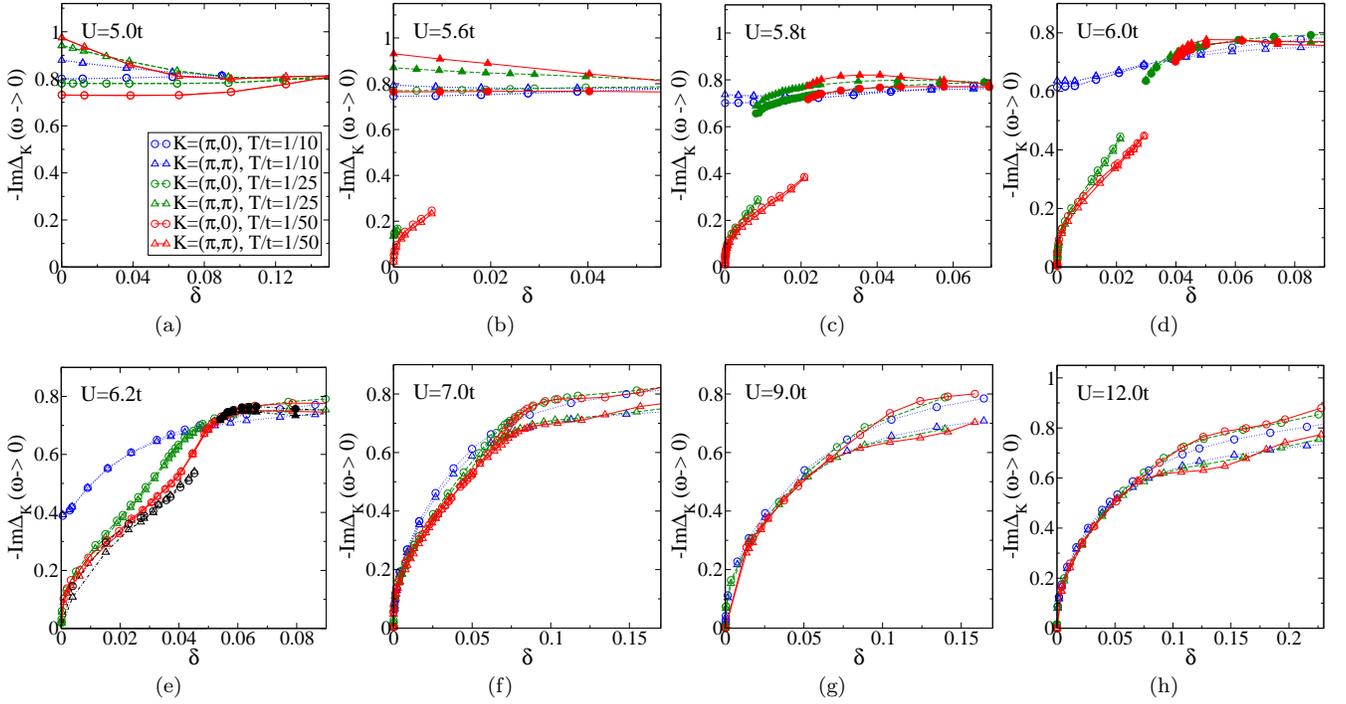

\centering
\subfigure[]{\label{fig13a}\includegraphics[width=0.24\linewidth]{fig13a.eps}}
\subfigure[]{\label{fig13b}\includegraphics[width=0.24\linewidth]{fig13b.eps}}
\subfigure[]{\label{fig13c}\includegraphics[width=0.24\linewidth]{fig13c.eps}}
\subfigure[]{\label{fig13d}\includegraphics[width=0.24\linewidth]{fig13d.eps}}

\subfigure[]{\label{fig13e}\includegraphics[width=0.24\linewidth]{fig13e.eps}}
\subfigure[]{\label{fig13f}\includegraphics[width=0.24\linewidth]{fig13f.eps}}
\subfigure[]{\label{fig13g}\includegraphics[width=0.24\linewidth]{fig13g.eps}}
\subfigure[]{\label{fig13h}\includegraphics[width=0.24\linewidth]{fig13h.eps}}
\caption{Extrapolated zero-frequency value of the imaginary part of the cluster hybridization function, -Im$\Delta_K(\omega\rightarrow 0)$, for cluster momentum $K=(\pi,0)$ (circles) and $K=(\pi,\pi)$ (triangles) as a function of doping, for different values of the interaction strength $U$. The data shown are for temperatures $T/t=1/10$ (blue dotted line) $1/25$ (green dashed line), $1/50$ (full red line), $1/100$ (black dot-dashed line). This observable measures the density of states of the conduction bath at the Fermi level averaged in the coarse grained cluster momentum region.
Note the weakly cluster momentum dependence in the underdoped phase, to be contrasted with the strong cluster momentum differentiation of both the cluster Green's function and self-energy.
}
\label{fig13}
\end{figure*}

Taking the perspective of the quantum impurity model immersed in a self-consistent bath is the best way to show this point.
In Fig.~\ref{fig12} we show the statistical weight of the main plaquette states as a function of doping and for several temperatures.
In Fig.~\ref{fig13} we show the extrapolated zero-frequency value of the imaginary part of the cluster hybridization function as a function of doping and for different temperatures.
The onset of the Mott insulating state ($\delta=0$ for $U>U_{\rm MIT}$) is signaled by a dramatic growth of the half filled singlet probability (black circles).
In the Mott state the half-filled singlet dominates and its probability increases with increasing interaction strength $U$.
The cluster hybridization function represents the effect of the bath in which the plaquette is immersed and with which electrons are exchanged. It is gapped at the Fermi level for all cluster momenta (i.e. Im$\Delta_K(\omega\rightarrow 0)\rightarrow 0$).
Together, the insulating character of the reservoir and the single dominant plaquette eigenstate indicate that the electrons of the lattice are localized and bound into short-range singlets due to superexchange interaction.

In the UD phase, the $2\times 2$ plaquette is still mostly occupied by the half-filled singlet (black squares), revealing that this phase evolves out of the parent Mott insulator.
However, doping the Mott state populates other plaquette eigenstates and the system begins to fluctuate between the half-filled singlet and other configurations.
Not surprisingly, the primary fluctuations introduced by doping are charge excitations, as shown by the increase of the weight of the doublet at $N=3$ (blue circle) with increasing doping.
In a smaller measure, the probability of the spin triplet excitations (red triangles) also grows with doping.
Interestingly, spin excitations (both of singlet and triplet type) show a larger temperature dependence than the charge excitations.
Note also that the probability of the half-filled singlet saturates below a characteristic temperature.
The previous sections have demonstrated the metallic, compressible character of the UD phase.
The fluctuations among various plaquette configurations are allowed by the self-consistent bath, which in this phase is metallic (Fig.~\ref{fig13}) and depends only weakly on the cluster momentum, consistent with singlets states within the plaquette that dominate the behavior of the system.

Our results thus suggest to consider the UD phase as a phase in which the doped-hole carriers propagate into a sea of spins bounded into a $N=4$ short-range singlet state (a similar interpretation for low doping phase of the $t-J$ model was suggested in Refs.~\onlinecite{hauleAVOIDED,hauleDOPING}).
We can therefore rationalize why this phase is coherent at low $T$, as revealed by the small scattering rate in Fig.~\ref{fig10}: as soon as the spins pair up into singlet pairs, the doped holes have no difficulty to delocalize coherently into the background of singlet states.

We now examine the OD phase.
Here there is no dominant plaquette eigenstate, and the electrons within the plaquette fluctuate among several states.
Electrons spend comparable time visiting the doublet at $N=4$, the half-filled singlet and triplet. Also important are other types of excitations.
At large doping, the charge excitations becomes important, so the probability of the doublet at $N=3$ is the largest.
As the transition is approached, the weight of these excitations decreases, while that of the singlet increases.
The temperature dependence of the plaquette eigenstates is weak and mostly confined to excitations of spins (both singlet and triplet).
To complete our survey of the OD phase we analyze the effect of the bath in which the plaquette is immersed, Fig.~\ref{fig13}.
We find that the cluster hybridization function displays metallic behavior and weak cluster momentum differentiation, especially as the finite-doping transition is approached.
These results for the OD phase support a qualitatively different picture compared with the UD phase.
At the level of the quantum cluster impurity problem, the OD phase is characterized by the electrons of the plaquette fluctuating among several configurations in a conducting bath. Here the bath plays a central role, since the electrons of the metallic bath screens the fluctuating moments via the Kondo effect.
The objects that delocalize in the lattice are both spin and charge excitations compatible with Landau quasiparticles that carry both spin and charge.
As a further support to this picture, we stress that at large doping, far from the transition, we recover the results of single-site DMFT, where the spatial fluctuations are absent and where the Kondo screening mechanism controls the physics of the model.

The central question then is: how can the system go from the OD phase to the UD phase?
At the level of the plaquette, the transition between the two phases results from the competition between singlet spin excitations and spin triplet plus charge excitations.
A simple way to quantify the increased importance of spin triplet-charge fluctuations at the expense of the singlet across the transition, is to monitor when the probability of the singlet state equals the probability of other plaquette states, i.e. when  $P_{s}=1/2$ (green solid line in Fig.~\ref{fig12}). This occurs roughly at the UD-OD transition.

The competing spin-charge excitation close to the transition was already reported in our recent study, Ref.~\onlinecite{sht}.
There, we emphasized the interplay between the doublet and the spin excitations.
Close to the critical coupling $U_{\rm MIT}$ (cf. Fig.~\ref{fig12}b,c,d,e) we linked the coexistence region to the crossing of the probability between the triplet and the doublet states. 
In the present study, which explores a larger region of interaction strength $U$ and temperatures, we rather highlight the competing singlet spin versus triplet-spin and charge fluctuations.
This aims to refine our previous study because the crossing $P_{s}=1/2$ tracks both the first-order transition at low temperatures and its high temperature crossover providing a more general framework for interpreting our results in the whole range of $U$ and $T$ explored.

We can now address the crucial question of the origin of large scattering rate close the the $\mu_{c1}(T,U)$ surface and its high temperature crossover analyzed in Sec.~\ref{Sec:Phases}.
The results of the plaquette eigenstates describe competing~\cite{hauleAVOIDED} superexchange and Kondo interactions that give rise to unquenched short-range magnetic fluctuations of the electrons within the plaquette.
This identifies the source of scattering close to the transition between the UD and OD phase.

The above description takes the point of view of the cluster coupled to a bath.
The cellular DMFT results have also a complementary interpretation in terms of the lattice.
In the OD phase the electrons on the lattice are coherent because they fill the available $k$ states as in a Fermi liquid~\cite{zimanBOOK}. At zero temperature, this quenches the entropy. Here the $k$ space picture is appropriate.
On the other hand, in the UD phase the electrons pair up into short-range singlets and doped charge carriers can move coherently into a background of singlet bonds. Here a real-space picture is more appropriate.
Starting at large doping and reducing $\delta$ towards the transition to the UD phase, the electrons reorganize their motion according to the strength of short-range magnetic spin correlation that increase as the Mott state at $\delta=0$ is approached.
Breaking of the electrons as entities carrying both spin and charge occurs at the finite doping transition. This releases entropy as confirmed in Sec.~\ref{Sec:Thermodynamics}.
Once the new excitations in the form of short range singlets are formed, and the system get settled into a magnetically stable configuration at low $T$, the system can acquire again coherence.
This perspective identifies the reorganization of the basic excitations of the system at the transition between the OD and UD phase as the source of scattering.
Such reorganization also occurs in some recent theoretical proposals that envision a fractionalisation of the electrons~\cite{fract1,fract2}. A different type of reorganization occurs in the charge $2e$ boson theory~\cite{PhillipsRMP:2010}.

\begin{figure*}[!ht]
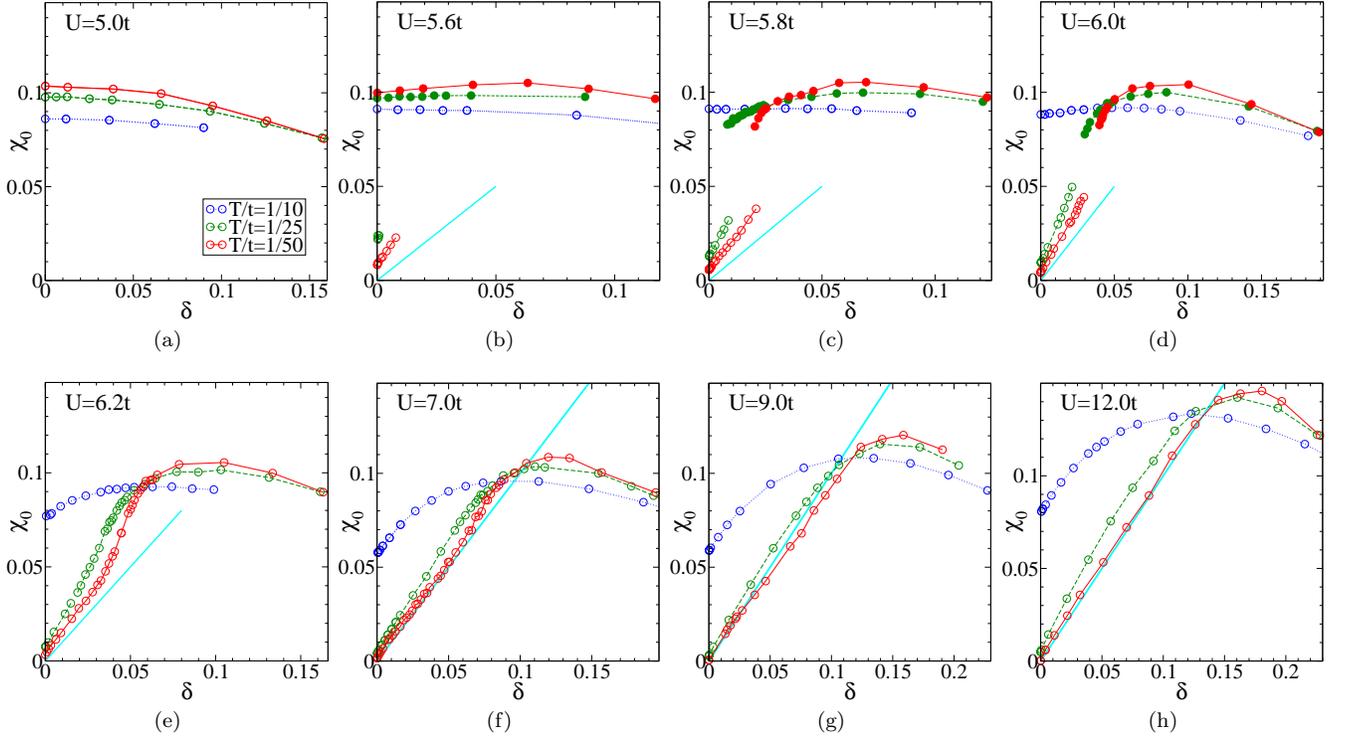

\centering
\subfigure[]{\label{fig14a}\includegraphics[width=0.24\linewidth]{fig14a.eps}}
\subfigure[]{\label{fig14b}\includegraphics[width=0.24\linewidth]{fig14b.eps}}
\subfigure[]{\label{fig14c}\includegraphics[width=0.24\linewidth]{fig14c.eps}}
\subfigure[]{\label{fig14d}\includegraphics[width=0.24\linewidth]{fig14d.eps}}

\subfigure[]{\label{fig14e}\includegraphics[width=0.24\linewidth]{fig14e.eps}}
\subfigure[]{\label{fig14f}\includegraphics[width=0.24\linewidth]{fig14f.eps}}
\subfigure[]{\label{fig14g}\includegraphics[width=0.24\linewidth]{fig14g.eps}}
\subfigure[]{\label{fig14h}\includegraphics[width=0.24\linewidth]{fig14h.eps}}
\caption{Local spin susceptibility $\chi_0$ versus $\delta$ for several values of $U$. The data shown are for temperatures $T/t=1/10,1/25, 1/50$ (blue dotted, green dashed, red solid line respectively). In the underdoped phase $\chi_0$ is approximately linear with doping, as highlighted by the solid cyan line, substantiating the idea that the doped carriers propagate among short-range singlets formed by the superexchange interaction.
}
\label{fig14}
\end{figure*}
Further support for our scenario follows from the the zero frequency cluster susceptibility $\chi_0=\int_{0}^{\beta} \langle S^c_z(\tau)S^c_z(0)\rangle d\tau$, where $S^c_z$ is the projection of the total spin along the z direction of the $2\times2$ plaquette, shown in Fig.~\ref{fig14} as a function of doping for several values of $U$ and $T$.
In the Mott insulating phase this quantity vanishes as temperature is lowered, revealing that the system is locked into $S=0$ singlet states due to the superexchange interaction.
This is in sharp contrast with single-site DMFT where the corresponding local spin susceptibility diverges in the Mott state.
As the doping is increased, $\chi_0$  grows approximate linearly with doping.
This characterizes the UD phase and reinforces the view of this region as a phase with hole-doped carriers propagating in a background of singlet states.
On the other hand, starting at large doping, a decrease of $\delta$ towards the transition has again the effect of slightly increasing $\chi_0$. The mechanism however is different: spin and charge excitations are Kondo screened by the conduction electrons of the bath. In a two-dimensional Fermi liquid with spherical Fermi surface, the spin susceptibility depends only on the density of states and is independent of doping in the non-interacting limit. Here, as the transition is approached with decreasing doping, $\chi_0$ increases slightly, then approximately saturates or slightly decreases, revealing the increased importance of the unquenched spin fluctuations in the plaquette as the transition is approached.
Notice also the opposite temperature behavior of $\chi_0$ within the two phases, again supporting the two different low lying elementary excitations in these two regions. In the UD phase in particular, the spin susceptibility decreases rapidly with temperature, in analogy with the spin-gap phenomenon found in high-temperature superconductors \cite{Alloul:1989,Warren:1989}.

\section{Discussion}
\label{Sec:Discussion}

Understanding the origin of the unusual behavior of copper oxide superconductors is a fascinating challenge for theorists and experimentalists alike.
Recent experiments are focusing on the normal state phase of these systems and are providing new clues on the critical behavior observed between the metallic state at high doping and the Mott insulator at zero doping~\cite{LouisREV}.
In an attempt to obtain a unified picture of these systems, we have mapped out the normal state phase diagram of the 2D Hubbard model, which is arguably the simplest model able to capture the essential physics of doped Mott insulator~\cite{andersonBOOK}.
Cellular DMFT on a $2\times2$ plaquette is our theoretical tool to elucidate the phase diagram, allowing us to consider both temporal and short-range spatial fluctuations.
Our results provide a unified low-temperature landscape where, at the lowest temperatures, the system evolves, as a function of the carrier concentration, from a Mott insulator at zero doping, through a coherent electron liquid in the UD region, and, with further doping, to a conventional metallic state in the OD region.
Our key finding is the first-order transition that takes place at low temperature between these two metallic phases at finite doping, which is a source of large scattering rate at the temperatures relevant for experiment.

In this section we further discuss the emerging phase diagram with a focus on the typical signatures and precursors of the first-order transition.
By tracking these indicators down to the Mott critical endpoint $U_{\rm MIT}$ we reveal that Mott physics is at the origin of this transition.
Subsequently we conjecture that the large scattering arising from this transition may provide an alternative to the quantum critical behavior surmised in the copper oxide superconductors\cite{Aeppli,Tallon}.

\subsection{Mott physics beyond half-filling}
\label{SubSec:dopings}

\begin{figure}[!ht]
\centering{\includegraphics[width=0.9\linewidth,clip=]{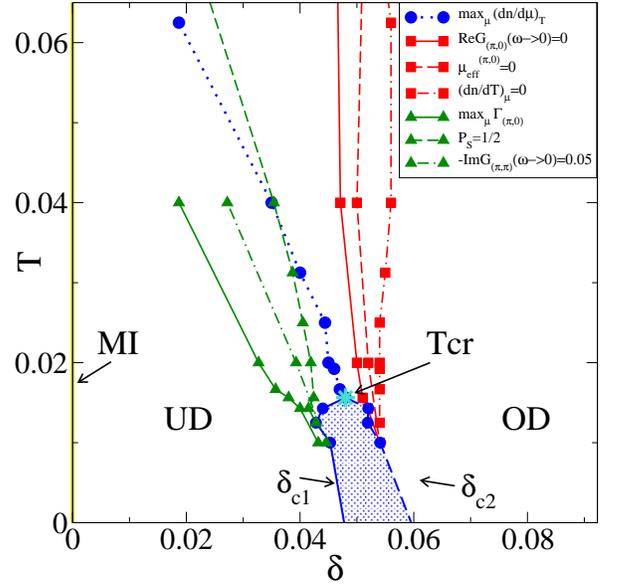}}
\caption{Phase diagram temperature $T$ versus doping $\delta$ for $U=6.2t$, representative of the regime $U>U_{\rm MIT}$. The star symbols denotes the second-order finite temperature critical point, with coordinates $(\delta,T)\approx (0.05,1/65)$.
Blue lines with circles identify the spinodals $\delta_{c1}$ (full line) and $\delta_{c2}$ (dashed line) where the UD and OD phases respectively disappear. Dotted blue line with circles denotes the high-temperature crossover of the transition measured by a maximum in the compressibility $(dn/d\mu)_{T}$.
The red lines with squares indicate the vanishing of Re$G_{(\pi,0)}(\omega\rightarrow 0)$ (solid line) and of $\mu_{\rm eff}^{(\pi,0)}$ (dashed line) and the crossing of the isotherms $n(\mu)$ (dot-dashed line) and they are early warning indicators of the spinodal $\delta_{c2}(T)$.
The green lines with triangles are precursors of the spinodal $\delta_{c1}(T)$. Solid line marks the maximum of the scattering rate $\Gamma_{(\pi,0)}=-Im\Sigma_{(\pi,0)}(\omega\rightarrow 0)$. The strong momentum differentiation found in Im$G_{K}(\omega\rightarrow 0)$ disappears at the dot-dashed line. When that differentiation occurs as a crossover, we defined the boundary by -Im$G_{(\pi,\pi)}(\omega\rightarrow 0)=0.05$. 
The dot-dashed line corresponds to the doping below which the probability of the half filled singlet state becomes larger than 1/2. Below the critical point the metallic state separates into two different metallic states, the UD and the OD phase, with a hysteretic change in the occupation (the jumps in the occupation define hatched region). The lines above the critical point are crossover lines. The Mott insulator (MI) exists only on the $\delta=0$ line.
}
\label{fig15}
\end{figure}

Fig.~\ref{fig15} shows the temperature-doping phase diagram for $U=6.2t$ representative of the regime $U>U_{\rm MIT}$ and summarizes our results for the doping induced transitions starting from the Mott insulator.
In the $T-\delta$ plane there are five regions of interest: Mott insulator, underdoped phase, hysteretic region, overdoped phase, and the non-Fermi liquid region arising from the finite-temperature crossover between the UD and OD phases.

{\it Mott insulator.} At zero doping the system is a Mott insulator, as revealed by a plateau in the occupation at the (odd) number $n=1$ as a function of $\mu$ (see Sec.~\ref{Sec:PhaseDiagram}).
Here the spins are bound into short-range singlets due to superexchange mechanism (see Sec.~\ref{Sec:Plaquette}).

{\it Underdoped phase.} Doping of the MI state proceeds gradually in certain cluster orbitals, with carriers going into the $(0,0)$, $(0,\pi)$ and  the degenerate $(\pi,0)$ orbital, but not in the $(\pi,\pi)$ one, that remains insulating.
This strong orbital differentiation is robust with respect to an increase of the size of the cluster immersed in the self-consistent medium~\cite{michelCFR}. 
Notice however that in CTQMC calculations this gradual doping in cluster orbitals is associated with a large orbital differentiation of the cluster self-energy, while exact diagonalization methods show the latter phenomenon without sign of the former, at least for a $2\times2$ plaquette~\cite{liebsch,civelliBreakup}. 
The basic excitations of this phase show coherent behavior, as indicated by the linear low frequency behavior of Im$\Sigma_{(\pi,0)}$ and Im$\Sigma_{(0,0)}$. However the system falls outside Fermi liquid theory since the $(\pi,\pi)$ orbital remains gapped in this phase suggesting a ``small'' Fermi surface compatible with the presence of arcs or pockets (see Sec.~\ref{Sec:Phases}).
Furthermore, the objects that delocalize through the lattice are holes propagating into a background of short-range singlet states, as suggested by the analysis of the plaquette eigenstates (see Sec.~\ref{Sec:Plaquette}).
The metallic character of this phase is revealed by the finite spectral weight at the Fermi level in the $(\pi,0)$ and $(0,0)$ orbitals and by the finite charge compressibility deduced from the $dn/d\mu|_T$ behavior.
Since the basic excitations of this phase are not the usual Landau quasiparticles, it will be interesting to explore the way these excitations carry heat and charge.

{\it Overdoped phase.} Upon further doping, there is a first-order transition to a phase that appears to be a conventional Fermi liquid phase.

{\it Coexistence region.} Below the characteristic temperature $T_{\rm cr}$, there is an hysteretic region in the $T-\mu$ plane where the OD and UD phases are both present. 
Therefore within this region a first-order transition takes place where the free energy of the UD and OD phase crosses. 
One of the two phases is metastable, but a computation of the free energy to tell which one is metastable would be prohibitive.
In the $T-\delta$ plane there is forbidden region delimited by the spinodals, which corresponds to the discontinuous jump of the occupation $n(\mu)$. 
If we were able to exclude the metastable phases, this forbidden region would be wider and delimited by the so-called binodals.

{\it Crossover region.} Above $T_{\rm cr}$, between the UD and the OD phase there is an incoherent phase signalled by a large scattering rate in the $(\pi,0)$ orbital.
This region has a funnel-like shape in the $T-\delta$ parameter space. It is noteworthy that the lower is $T_{\rm cr}$, the higher is its extension in temperature.
Future work should better clarify the thermodynamic critical behavior of this region as well as its signatures on transport properties. We note nevertheless that there is a maximum in entropy as a function of doping close to the boundary of the OD phase, see.~\ref{Sec:Thermodynamics}. This occurs because the two phases 
have different electronic excitations. At the first order transition, or just above, the electronic excitations of the system are thus reorganized leading to a large entropy.

Let us now review the phase boundaries.
As it appears from Fig.~\ref{fig15}, the phase diagram is dominated by the critical point at the characteristic temperature $T_{\rm cr}$ (star symbol).
Below this critical point the metallic state separates into two distinct metallic states: the UD state to the left of the critical point and the OD state to the right.
The spinodal lines $\delta_{c1}$ and $\delta_{c2}$ emanate from the critical point (solid and dashed blue curves with circles) and are defined as the loci in the phase diagram where the occupation $n=1-\delta$ undergoes a sudden drop and an upward jump respectively.
Hence, the region between $\delta_{c1}$ and $\delta_{c2}$ is forbidden and translates into a coexistence of phases with different densities in the $T-\mu$ plane with hysteretic behavior clearly apparent in the $n(\mu)$ curves (see Sec.~\ref{Sec:Transition}).
Above the critical point at $T_{\rm cr}$ there is a characteristic crossover line, denoted by the dotted blue line with circles, and defined by the peak in the charge compressibility $dn/d\mu|_{T}$.

To detect directly the first-order transition by looking for hysteresis in some observables, like the occupation, requires access to low temperatures and a careful scan of phase space.
Since $T_{\rm cr}$ rapidly moves to low $T$ as $U$ increases
it is very helpful to find some precursory signals that allow to infer such a transition.

The first indicators that the system is approaching $T_{\rm cr}$ is the above-mentioned crossover line defined by the maximum in the compressibility and the associated critical slowing down in our CTQMC calculations close to that line (see Sec.~\ref{Sec:Transition}).
Interestingly, we also found precursors of the spinodal lines $\delta_{c1}$ and $\delta_{c2}$, where the UD and OD phases cease, respectively, to exist.
Signs that the model is approaching the spinodal $\delta_{c2}$ include the vanishing of Re$G_{(\pi,0)}(\omega\rightarrow 0)$ and of the renormalized chemical potential $\mu_{\rm eff}^{(\pi,0)}$ and the crossing of the isotherms $n(\mu)$ (solid, dashed and dot-dashed red curves with squares in Fig.~\ref{fig15}).
These indicators originate at low temperatures close to the $\delta_{c2}$ line and persist for temperatures higher than $T_{\rm cr}$.
The first two phenomena reflect a low frequency particle-hole symmetry of the orbital $(\pi,0)$ at this critical line.

We now compile the characteristic precursors of the spinodal $\delta_{c1}$, shown as green curves with triangles in Fig.~\ref{fig15}.
The indicators includes the maximum of the scattering rate of the $(\pi,0)$ orbital $\Gamma=-Im\Sigma_{(\pi,0)}(\omega\rightarrow 0)$ (solid line), the strong orbital differentiation in the density of states at the Fermi level, measured by -Im$G_{K}(\omega\rightarrow 0)$ (dot-dashed curve) and the line where the probability of the spin-singlet plaquette state that dominates at half-filling exceeds $1/2$.
Below $T_{\rm cr}$ these phenomena take place close to the spinodal line $\delta_{c1}$, therefore directly signalling the transition.
Above $T_{\rm cr}$, they may serve as indicators that the transition is approaching. We note that these precursory signals can be detected at rather large temperatures.
In addition, we have verified that as $T_{\rm cr}$ decreases with increasing $U$, these indicators can actually be detected at higher and higher temperatures, a definite advantage when it comes to tracking the position of $T_{\rm cr}$ when it becomes very low.
This behavior is quite surprising and calls for a better understanding of the high $T$ crossover above the transition. 
It seems that when $T_{\rm cr}$ is very low, it effectively acts as a quantum critical point. 
Quantum criticality influences the physics at rather high temperature\cite{SachdevBOOK:2001,Kopp:2005,RoschRMP:2007,Roy:2008,vlad}.

\begin{figure}[!ht]
\centering{\includegraphics[width=0.9\linewidth,clip=]{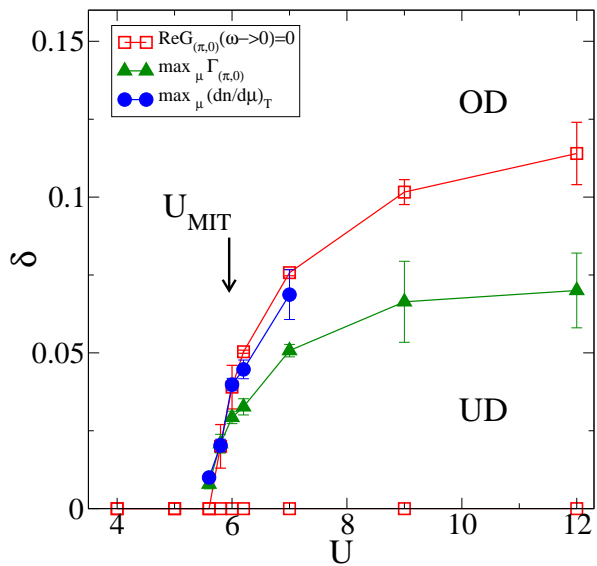}}
\caption{Characteristic dopings versus $U$ that signal the UD to OD transition, measured at the temperature $T/t=1/50$. The blue line with circles denotes the maximum of the compressibility $(dn/d\mu)_T$. The red full line with squares indicates the vanishing of Re$G_{(\pi,0)}(\omega\rightarrow 0)$. Green line with triangles denotes the maximum of the scattering rate $\Gamma_{(\pi,0)}=$-Im$\Sigma_{(\pi,0)}(\omega\rightarrow 0)$. Data points are displayed for the constant temperature $T/t=1/50$ and thus they measure the features of the three-dimensional phase diagram {\it at that temperature}. The overall maximum (as a function of $U$, $T$ and $\delta$) of the compressibility follows the critical line $T_{\rm cr}$ that originates at $U_{\rm MIT}$, with coordinates $(U,\delta,T)\approx(5.95,0,1/12)$ indicated by the vertical arrow. The overall maximum as a function of $U$, $T$ and $\delta$ of the scattering rate $\Gamma_{(\pi,0)}$ follows the $\delta_{\rm c1}(U,T)$ surface and peaks at $U_{\rm MIT}$. The zero of the extrapolated value of real part of the $(\pi,0)$ cluster Green's function at $\omega=0$ bifurcates at the line $U_{c2}(T)$.
}
\label{fig16}
\end{figure}

We now show that all the above indicators are manifestations of Mott physics.
If we follow them as a function of the correlation strength $U$ (at a given temperature $T/t=1/50$), we obtain the results shown in the $\delta-U$ plot of Fig.~\ref{fig16}.
The critical line $T_{\rm cr}$ (blue curve with circles) originates at the Mott endpoint $U_{\rm MIT}$ with coordinates $(U/t\approx5.95,\delta=0)$ and moves to progressively larger doping as $U$ is increased, as discussed in detail in Sec.~\ref{Sec:Transition}.
The line tracking the $U$ evolution of the maximum in scattering rate of the $(\pi,0)$ orbital (green curve with triangles) originates at the spinodal $U_{c1}$ at zero doping  and departs from zero doping with increasing $U$. 
Due to particle-hole symmetry, the indicator Re$G_{(\pi,0)}(\omega\rightarrow 0)=0$ (red line with squares) is present at zero doping for all values of $U$ but, in addition, at the spinodal $U_{c2}$ the curve bifurcates and a new branch splits away from zero doping as $U$ increases.

The fact that all signatures of the finite doping first-order transition can be traced back to the Mott critical point reveals that the first-order transition between the UD and the OD phase is associated with Mott physics.
This in turn implies that the signature of the Mott transition in the two-dimensional Hubbard model, normally expected to occur in close proximity to half-filling, extends way beyond half filling.

This surprising conclusion~\cite{sht} emerges naturally from the detailed mapping of the 3D normal state phase diagram.
It is indeed quite difficult to derive such a picture from the analysis of the doping driven transition at large $U$ only.
Studies of this type may detect a critical doping and the strong orbital differentiation approaching the Mott insulator.
It would be hard, though, to find that the detected critical doping is indeed a high temperature signature of a finite $T$ first-order transition, because of the low energy scales involved (i.e. the critical line $T_{\rm cr}$ moves rapidly to progressively lower temperature as $U$ increases).
It would also be hard to make an unambiguous link between that transition at finite doping and the Mott endpoint at zero doping.

The map of the whole normal-state phase diagram, along with our discovery of a first-order transition originating at the half-filled Mott critical point and moving progressively away from half-filling as $U$ increases, should help provide a unified picture of a host of previously known results and call for their reinterpretation as precursory signals of the transition we have revealed. 
In this respect, and accounting for the differences in methods, the pseudogap - Fermi liquid transition reported for $U$ larger that the Mott endpoint in Ref.~\onlinecite{liebsch} as well as the crossover observed at large doping in Refs.~\onlinecite{jarrellNFL,jarrell:2009,khatami} are likely to be the same as those observed here.
Nevertheless, our analysis suggests that the quantum critical point hypothesized in Refs.~\onlinecite{Aeppli,Tallon} is instead a very low {\it but finite T} critical point that originates from the influence of Mott physics away from half filling.
We cannot strictly rule out that the critical line that ends our first-order surface becomes a quantum critical line or point at some $U$ (see theoretical investigations by Imada in Refs.~\onlinecite{imadaCR1,imadaCR2}), however this hypothesis is disfavored in our calculations by the fact that the same precursory signs of the transition occur for all the values of $U>U_{\rm MIT}$ considered, without apparent qualitative change (see Fig.~\ref{fig16}). The same argument holds to disfavor the hypothesis of a tricritical point at some $U$ in the phase diagram.  
Several earlier studies have also shown that the strong orbital differentiation phenomenon is a robust feature of cluster DMFT solutions of the two-dimensional Hubbard model~\cite{michelCFR}. 
But it was unclear what is the origin of the strong momentum differentiation and of its disappearance with doping. 
This phenomenon finds a natural explanation in our phase diagram since it marks the end of the UD phase.

Even if more investigations should be done to clarify both the crossover behavior above $T_{\rm cr}$ and the zero temperature limit of our phase diagram, our study reveals that a first-order transition inhabits and controls much of the physics of the normal-state phase diagram of the two-dimensional Hubbard model.

\subsection{Proposal for critical behavior}
\label{SubSec:criticality}

We now turn to the relevance of our results for the physics of the copper oxide superconductors~\cite{normanADV,pepin,bonn,lee,normanPhy}.
The difficulties to achieve a unified picture are both experimental and theoretical.
From the theory side, they arise mainly because the phase diagram of the cuprates defies three paradigms of condensed matter physics: band theory fails to explain the Mott insulator at zero doping, Bardeen-Cooper-Schrieffer description does not describe the unusual superconductivity and the Landau Fermi liquid theory does not capture the metal above $T_c$ and the pseudogap phase close to the Mott insulator.
Therefore, understanding the evolution of these systems as a function of doping remains a challenging issue.
Since the parent compounds are Mott insulators, this calls for an understanding of the doping-driven MIT arising from doping the Mott insulator~\cite{ift}.
Models of this phenomenon rely on simple Hamiltonians such as the Hubbard model.
Here the aim is to ascertain to what extent this minimal model is able to capture the essence of the complex physical behavior of the copper oxide superconductors.
We mention that in this context the optical lattices of cold atoms have opened up new perspectives to test these assumptions~\cite{jzHM,lewenstein,bdzRMP} and determine the phase diagram of basic Hamiltonians of condensed matter physics~\cite{greiner,Jordens:2008,EckardtPRA}

From a methodological viewpoint, dynamical mean-field theory approaches allow to follow different phases as a function of model parameters.
In this  paper we concentrated on the normal state, by studying the cellular DMFT paramagnetic solution of the two-dimensional Hubbard model.
Thus the main experimental relevance of our work consists in elucidating how the proximity to the Mott insulator affects the normal state.

Our key finding, namely the first-order transition between two metals, suggests the conjecture that the quantum critical behavior surmised in high-temperature superconductors \cite{Aeppli,Tallon} originates from a constant $U$ cut of the finite-doping critical line ending the first-order surface between two types of metals. 
That transition occurs only when $U$ is larger than the critical value $U_{c1}$ for the Mott transition and moves to progressively higher doping as $U$ increases, reaching $\delta\approx0.12$ at $U=12t$, which corresponds to the doping that maximizes the superconducting transition temperature~\cite{hauleDOPING,hauleAVOIDED}. 
This suggests that our first-order transition lies deep below the superconducting dome (cf. Refs.~\onlinecite{hauleDOPING,hauleAVOIDED,khatami}). 
In addition, at large $U$ the critical point $T_{\rm cr}$ is at extremely low temperature and the smaller $T_{\rm cr}$ the farther its influence, in the form of large scattering rate, extends at finite temperature\cite{SachdevBOOK:2001,Kopp:2005,RoschRMP:2007,Roy:2008,vlad}.
Our contribution is to pin down the large scattering rate to the influence of the Mott phenomenon far away from half filling: the first-order transition is the source of anomalous scattering, which displays a funnel shape with increasing temperature. 
Such funnel shape scattering separates underdoped from overdoped phases
above the maximum of the superconducting dome in high temperature
superconductors. 
Therefore the UD and OD phases, that we defined with
respect to our metal-metal first order transition, should also correspond to
the underdoped and overdoped regions normally defined in high temperature superconductors. 
Given the differences in methods, the crossover that we found between UD and OD phases is also the one detected for $U=8t$ in Refs.~\onlinecite{jarrellNFL,khatami}.

Further work is planned to make the correspondence clearer, but at present at the qualitative level several of our results are similar to those found in experiment. First, anomalous scattering takes place near a critical doping.
Contrary to other approaches~\cite{siREV}, the source of this scattering is not a quantum critical point between two ordered phases, nor the proximity to an ordered phase, but a finite temperature crossover at finite doping coming from Mott physics at work far from half filling. 
Second, above the critical point, the thermal expansion coefficient $(\partial n/\partial T)_{\mu}$ vanishes and equivalently, the entropy has an extremum $(\partial S/\partial \mu)_{T}=0$ \cite{khatami}. 
Using the Kelvin formula \cite{Shastry:2009}, this suggests that the thermopower changes sign at finite doping~\cite{shila}, consistent experiment \cite{Honma:2008}. Finally, as observed in the pseudogap phase \cite{Alloul:1989,Warren:1989}, there is a strong decrease with temperature of the spin susceptibility in our underdoped phase.

\section{Conclusion}

The complete normal-state phase diagram of the two-dimensional Hubbard model as a function of temperature, doping and interaction strength reveals a surface of first-order transition that ends on a critical line and that can separate two different types of metals. 
That transition at finite doping between the underdoped and overdoped phase is unexpectedly connected to Mott physics, since it originates at the half-filled Mott endpoint and moves progressively away from half-filling as the interaction strength increases. 
The critical line rapidly moves to low temperature as interaction strength is increased beyond the critical value for the Mott transition, and it leaves finite-temperature signatures similar to those that have suggested the existence of a quantum critical point near optimal doping in high-temperature superconductors. Hence we propose that Mott physics leads to a very low temperature critical point separating two metals and that this can be the source of anomalous normal state behavior near optimal doping in high-temperature superconductors. Even though this critical point cannot be accessed because of intervening long-range order, it controls the physics at sufficiently high temperature in the same way that a zero-temperature Fermi-liquid ground state controls the normal state Fermi liquid despite the fact that the true ground state may have long-range order.

Our proposal is clearly in its infancy and further work, already in progress, should reveal the precise nature of the criticality along with the relationship between the pseudogap phase and the UD region of our phase diagram.
In addition, the study of some ordered states, like the antiferromagnetic and superconducting ones, should provide useful insight to complete the cellular DMFT picture of the Mott transition.
Nevertheless, our discovery sheds new light on the phase diagram of the 2D Hubbard model and highlights that Mott physics can play a pivotal role even far away from half filling.

\section*{ACKNOWLEDGMENTS}
We acknowledge Steve Allen for invaluable technical help and Patrick S\'emon and David S\'en\'echal for useful suggestions and discussions. This work was partially supported by FQRNT, by the Tier I Canada Research Chair Program (A.-M. S. T.), and by NSF DMR-0746395 (K. H.). Computational resources were provided by CFI, MELS, the RQCHP, and Compute Canada.


\end{document}